# Mechanistic Insight to the Chemical Treatments of Monolayer Transition Metal Disulfides for Photoluminescence Enhancement

*Zhaojun Li[1,2], Hope Bretscher[1], Yunwei Zhang[1], Géraud Delport[1], James Xiao[1], Alpha Lee[1], Samuel D. Stranks[1,3], and Akshay Rao[1]**

[1]Cavendish Laboratory, University of Cambridge, JJ Thomson Avenue, CB3 0HE, Cambridge, United Kingdom

E-mail: ar525@cam.ac.uk

[2]Molecular and Condensed Matter Physics, Department of Physics and Astronomy, Uppsala University, 75120 Uppsala, Sweden

[3]Department of Chemical Engineering & Biotechnology, University of Cambridge, Philippa Fawcett Drive, CB3 0AS, Cambridge, United Kingdom

Keywords: chemical treatments, transition metal disulfides, photoluminescence, trion suppression, p-doping

## Abstract

There is a growing interest in obtaining high quality monolayer transition metal disulfides (TMDSs) for optoelectronic device applications. Surface chemical treatments using a range of chemicals on monolayer TMDSs have proven effective to improve their photoluminescence (PL) yield. However, the underlying mechanism for PL enhancement by these treatments is not clear, which prevents a rational design of passivation strategies. In this work, a simple and effective approach to significantly enhance PL of TMDSs is demonstrated by using a family of cation donors, which we show to be much more effective than commonly used p-dopants which achieve PL enhancement through electron transfer. We develop a detailed mechanistic picture for the action of these cation donors and demonstrate that one of them, Li-TFSI (bistriflimide), enhances the PL of both $MoS_2$ and $WS_2$ to a level double that compared to the widely discussed and currently best performing “super acid” H-TFSI treatment. In addition, the ionic salts used in chemical treatments are compatible with a range of greener solvents and are easier to handle than super-acids, which provides the possibility of directly treating TMDSs during device fabrication. This work sets up rational selection rules for ionic chemicals to passivate TMDSs and increases the potential of TMDSs in practical optoelectronic applications.

The discovery of 2D materials based on semiconducting transition metal disulfides (TMDSs), with the chemical structure $MS_2$ (M=Mo, W), has opened up new interesting possibilities in optoelectronic devices, as monolayer TMDSs possess direct bandgaps with absorption in the visible spectral region, as well as other excellent properties well suited for optoelectronic applications, like high extinction coefficients due to the strong excitonic effects, exceptional mechanical properties, and chemical and thermal stability.[1–4] Nevertheless, monolayer TMDSs often exhibit poor photoluminescence quantum yields (PLQYs), which is the key figure of merit for optoelectronic devices.[5] Atomic vacancies, such as sulphur vacancies, which lead to trapping and non-radiative decay are thought to be the primary defects in these materials.[6,7] In addition, trion formation, which occurs easily in these materials, leads to non-radiative recombination and quenched photoluminescence (PL).[8–10] This is especially problematic since as-prepared TMDSs are often doped.[11,12]

To overcome these problems, there has been a large effort to develop chemical passivation strategies for TMDSs.[13–15] Chemical passivation by completing the dangling bonds has been widely used in silicon solar cells to improve the device performance.[16,17] For TMDS materials, the controlled physisorption of small molecules on the TMDS surface is reported to be a viable approach to tune their optical and electronic properties, but the increase in PLQY with these treatments is modest[11,18,19] In contrast, the use of 'acid treatment' with the super-acid trifluoromethanesulfonimide (H-TFSI) has been shown to greatly improve PLQY, up to 200 fold.[20] The mechanism for this improvement is still under debate, but has been suggested to involve the reduction of n-doping and trion formation, thus leading to increases in radiative recombination.[9,21–24] Despite these known treatments, the search for new passivating chemical treatments continues. Importantly, the harsh nature of the H-TFSI super-acid limits its application in optoelectronic devices, where it can cause damage to both the TMDS material and contacts. Here, we introduce a mild chemical treatment that uses ionic salt compatible with a diverse range of green solvents, performed under ambient condition. We demonstrate that bis(trifluoromethane)sulfonimide lithium salt (Li-TFSI) treatment yields larger PL improvements than the H-TFSI treatment, as well as greatly improved exciton diffusion compared to pristine or H-TFSI treated samples. By systematically studying the PL enhancement of TMDSs caused by different ionic chemicals, as well as widely used small molecule p-dopants, we provide mechanistic insight into the roles played by the cation and counter anion during chemical treatments. In addition, we show that the strong PL improvement is caused by cation adsorption on the TMDS surface instead of charge transfer to molecular p-dopants.

## Results and Discussion

The structures of all the chemical treatment agents used in this study are illustrated in Fig. 1a. The chemical treatments were achieved by immersing the monolayers in concentrated solutions of the investigated chemicals (0.02 M) for 40 mins. Because the PL of $MoS_2$ increases while increasing concentration of H-TFSI (Fig. S1), we compare all treatments with a fixed concentration of 0.02 M. Fig. 1b demonstrates the general PL enhancements on $WS_2$ with different chemical treatments. Apart from the H-TFSI super-acid reported previously, we find that a range of TFSI based ionic salts lead to PL enhancements to varying degrees. Interestingly, calcium (II) bis(trifluoromethanesulfonimide) ($Ca(TFSI)_2$) and Li-TFSI greatly improve PL, beyond what can be achieved *via* H-TFSI and other chemical treatments. It is also worth noting that compared to H-TFSI, which needs to be dissolved in dichloroethane (DCE) and has to be handled in the glovebox due to being extremely hygroscopic, ionic salts such as Li-TFSI and $Ca(TFSI)_2$ function in various milder solvents like acetonitrile, isopropanol and methanol, and can be easily handled in ambient atmosphere.

We start by comparing the improvement in PL intensity *via* treatment with H-TFSI and Li-TFSI. Representative PL spectra for pristine, H-TFSI and Li-TFSI treated monolayers $MoS_2$ and $WS_2$ are shown in Fig. 1c and 1d.[25] The PL of pristine $MoS_2$ and $WS_2$ has large contributions from trions with peak emission at 663 nm and 626 nm, respectively (see SI for spectral deconvolution). After chemical treatments, the PL is greatly enhanced and the peak position blueshifts for both $MoS_2$ and $WS_2$ due to the suppression of trions.[26] In addition, the PL enhancement yielded by the Li-TFSI treatment almost doubles that of the H-TFSI treatment, for both $MoS_2$ and $WS_2$ with an exciton emission peak at 659 nm and 617 nm, respectively. The blueshift in peak position is accompanied by a more uniform emission profile in both Li-TFSI and H-TFSI treated $MoS_2$ and $WS_2$. The distribution of the peak emission wavelength narrows with these treatments, as shown in scatter plots of the peak PL counts versus emission peak position acquired from PL spatial maps (Fig. S2). Other chemical treatments outlined in Fig. 1 exhibit various PL increase and peak position blueshift shown in Fig. S2 – S6. We will focus on Li-TFSI, as an example of a TFSI based ionic salt, in the discussion to follow, as Li-TFSI results in the highest PL enhancements in $MoS_2$ and $WS_2$ (see discussion on the other ionic salts $M_1$-TFSI ($M_1$ = Na and K) and $M_2(TFSI)_2$ ($M_2$ = Mg, Ca and Cu) in SI).

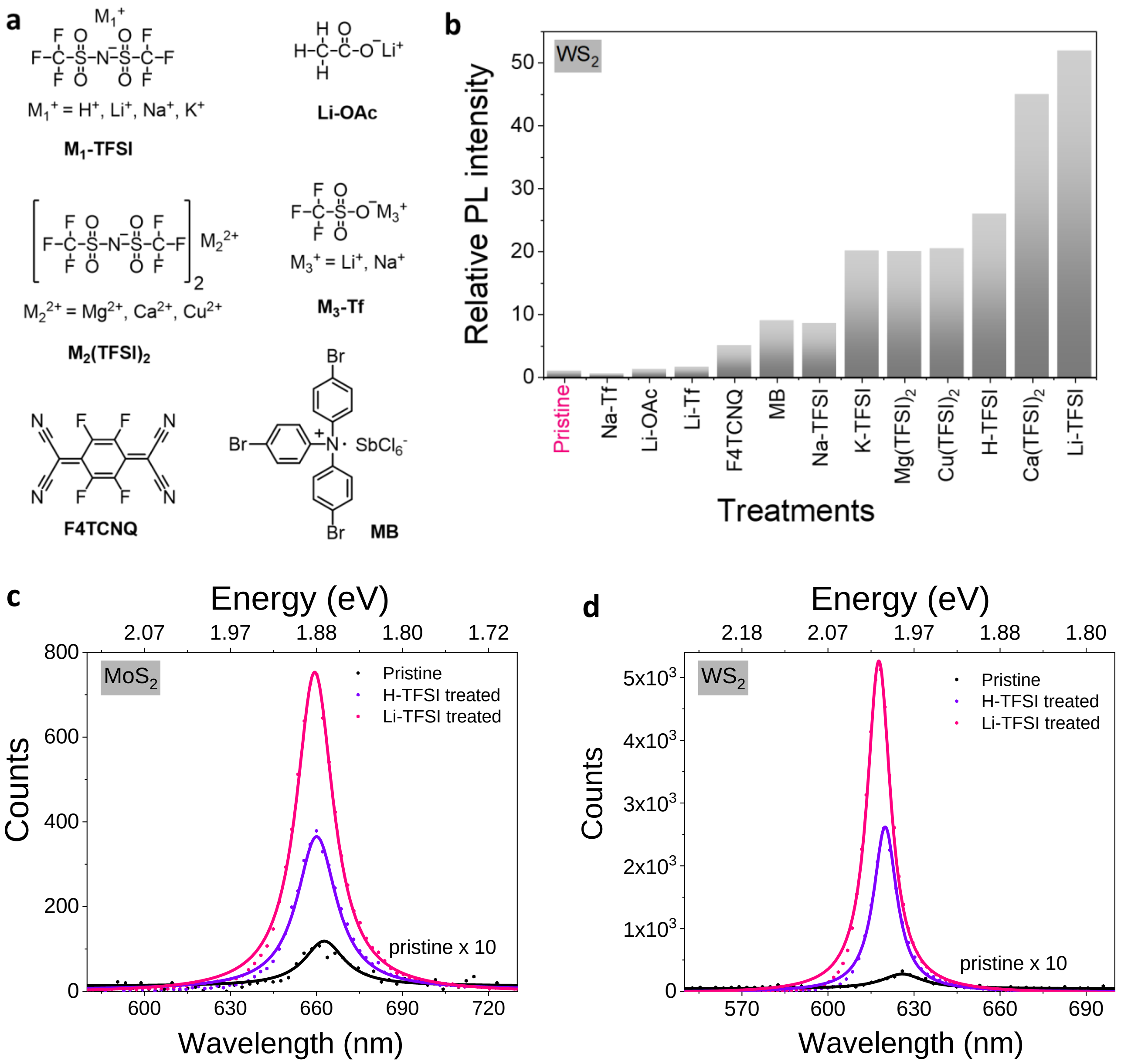

**Fig. 1. Studied chemicals and their steady-state photoluminescence (PL) enhancement on $MoS_2$ and $WS_2$. a** Structures of all the chemicals for the treatments. **b** General illustration of PL intensity enhancements on $WS_2$ with different chemical treatments compared to pristine sample (The PL intensity of pristine sample is normalized to 1). **c** Representative PL spectra for pristine, H-TFSI and Li-TFSI treated monolayer $MoS_2$. **d** Representative PL spectra for pristine, H-TFSI and Li-TFSI monolayer $WS_2$.

To study the mechanism of the PL enhancement, we carried out Raman spectroscopy on pristine H-TFSI, and Li-TFSI treated monolayer $MoS_2$ samples. A multi-peak Lorentzian fitting is performed on each spectrum to extract the chemical treatment dependent shift of the $MoS_2$ Raman peaks, as shown in Fig. 2. The in-plane $E_{2g}^1$ mode at 388 $cm^{-1}$ of pristine $MoS_2$ is associated with opposite vibration of two S atoms with respect to the Mo atom while the $A_{1g}$ Raman mode at 405 $cm^{-1}$ of pristine $MoS_2$ results from out-of-plane vibration of S atoms in opposite directions and is sensitive to doping-induced electron density.[27] The second order Raman resonance 2LA mode at 442 $cm^{-1}$ involving longitudinal acoustic phonons is assigned to in-plane collective movements of the atoms in the lattice.[28] After the treatment, both $A_{1g}$ and 2LA modes are blueshifted whereas the $E_{2g}^1$ mode is not affected. This is attributed to the weaker electron-phonon coupling caused by adsorption of the cations. Interestingly, a new Raman mode at ~ 468 $cm^{-1}$ emerges after H-TFSI and Li-TFSI treatments. This is assigned to the $A_{2u}$ mode, which is Raman-silent due to the reflection symmetry in pristine $MoS_2$.[29] The results imply that $H^+$ and $Li^+$ ions can be adsorbed at the surface of $MoS_2$, perturbing the crystal lattice and activating the previously silent $A_{2u}$ mode. As the PL enhancement is significantly greater for H-TFSI and Li-TFSI than p-doping small molecules, it suggests that PL modulation strength of the ionic chemicals might be determined by the interaction between the cation and the TMDS surface. This is in contrast to the common assumption that PL enhancement by surface chemical treatment is due to molecular p-dopant induced electron transfer.[9,30]

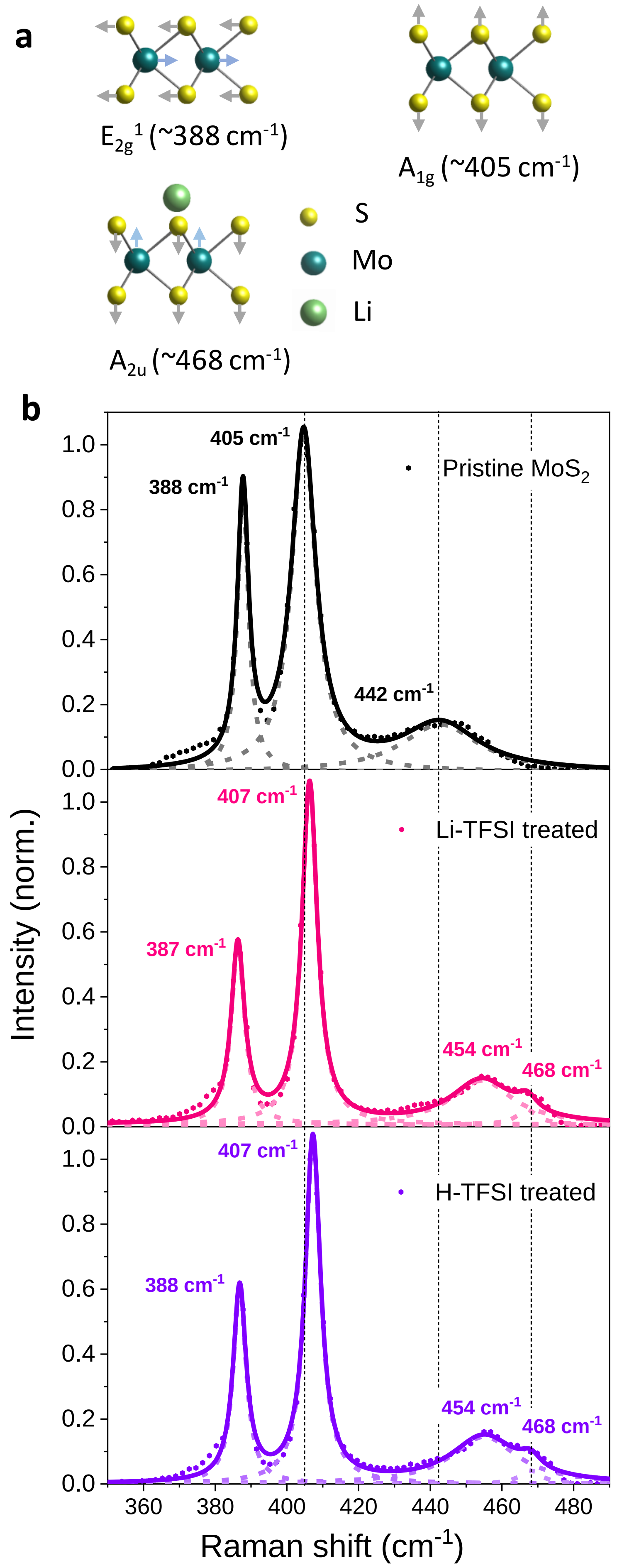

**Fig. 2 Raman spectra of chemically treated $MoS_2$ monolayers. a** Side views of Raman modes. **b** Raman spectra of pristine, H-TFSI-treated, and Li-TFSI-treated, and monolayer $MoS_2$. The decomposed Lorentzian peak fitting of each spectrum is presented as a short dashed line and the cumulative fitting is presented as a solid line. The positions of $A_{1g}$ and 2LA mode of pristine $MoS_2$ as well as $A_{2u}$ mode of $MoS_2$ with adatom (Li for example) are also illustrated in each spectrum for direct comparison.

To test our hypothesis of the importance of cation, the PL enhancements of $MoS_2$ and $WS_2$ treated with two common molecular p-dopants tris(4-bromophenyl)ammoniumyl hexachloroantimonate ("Magic Blue," MB) and 2,3,5,6-Tetrafluoro-7,7,8,8-tetracyanoquinodimethane (F4TCNQ) were investigated.[11,19,31] As depicted in Fig. S7, while both MB and F4TCNQ increased the PL of $MoS_2$ slightly, the enhancement is negligible in contrast to $M_1$-TFSI ($M_1$=H, Li, Na and K) and $M_2(TFSI)_2$ ($M_2$=Mg, Ca, and Cu) treatments. Moreover, there is a clear trion contribution from the emission of MB-treated $MoS_2$, and the PL of F4TCNQ-treated $MoS_2$ is too weak to obtain an accurate fitting. As illustrated in Fig. S8, both $A_{1g}$ and 2LA Raman modes of MB-treated and F4TCNQ-treated $MoS_2$ are slightly blueshifted due to the p-doping effect.[31] However, the shift is smaller compared to Li-TFSI-treated $MoS_2$, and there is no appearance of the $A_{2u}$ mode. This comparison strongly supports our hypothesis that PL enhancement of chemical-treated TMDSs is attributed to stable cation adsorption instead of electron transfer induced by molecular p-doping. This is further supported by PL enhancement of MB and F4TCNQ-treated $WS_2$ (See SI, Fig. S6, for detailed discussion). This stable cation adsorption effectively supresses trion formation in these materials.

Surface-sensitive X-ray photoelectron spectroscopy (XPS) measurements were also carried out on pristine, H-TFSI-treated, and Li-TFSI-treated $MoS_2$ samples to investigate the chemical treatment mechanism. As depicted in Fig. S9, the F 1s and Li 1s core levels show clear signatures of Li-TFSI adsorption on the sample surface.[32] Moreover, there is no observable change in oxidation state or bonding property according to the Mo 3d core levels which represent the Mo(IV) species.[33–35] Thus the peak at 169 eV in S 2p core levels is assigned to the TFSI anion instead of new oxidation state formation during the Li-TFSI treatment.[32] Though the S 2s peak, S 2p doublet peaks, and Mo 3d doublet peaks appear to shift towards higher oxidation states after H-TFSI and Li-TFSI treatments, the low sensitivity of the instrument prevents any interpretation of these effects.

To further test the above hypothesis, we investigate the stability of H and Li atom adsorbed at various types of adsorption sites in monolayer $MoS_2$ and $WS_2$ *via* density functional theory (DFT) simulations of the formation energy.[36] The formation energies of adatoms at sulphur vacancy sites ($E^{Sv}$), on top of sulphur ($E^{sf}$(S)), and on top of molybdenum ($E^{sf}$(Mo)) of $MoS_2$ are summarized in Table 1. The corresponding formation energy of cation-adsorbed $WS_2$, as well as the bond energy between cations and TFSI anion is listed in Table S1 (see SI for detail). The calculated results show that adsorption on both sulphur vacancy sites and on the top of surfaces of $MoS_2$ are thermodynamically stable with negative formation energies, but that the sulphur vacancy site rather than the surfaces of TMDSs is the most favourable adsorption location for all adatoms. In general, the adsorptions of Li adatom are energetically more favourable at surface sites ($E^{sf}$) compared to H adatom. However, H adsorption energy at sulphur vacancy site ($E^{Sv}{}_{H}$) is slightly more stable than Li ($E^{Sv}{}_{Li}$). Considering that the material has more available adsorption sites at the surface than sulphur vacancy, we believe that with chemical treatments, the concentration of Li adatoms on $MoS_2$ is higher than that of H adatom, due to the availability of locations for adsorption. This supports our assumption that the trion formation will be strongly suppressed by a higher adsorption of cation, leading to superior PL enhancement of TMDSs with Li-TFSI treatment.

**Table 1. DFT simulation of H and Li adatoms formation energies and the configurations on the different positions of monolayer $MoS_2$.**

| **Adatom** | $E^{Sv}$ (eV) | $E^{sf}$ (S) (eV) | $E^{sf}$ (Mo) (eV) |
|---|---|---|---|
| **H** | 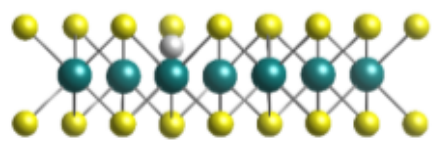 -2.41 | 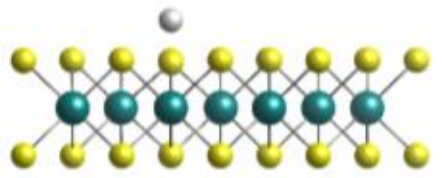 -0.41 | 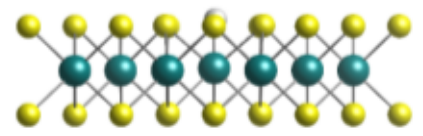 -0.016 |
| **Li** | 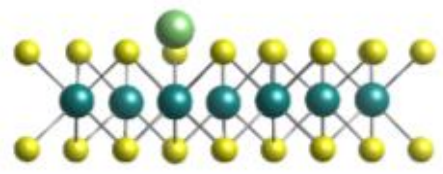 -2.39 | 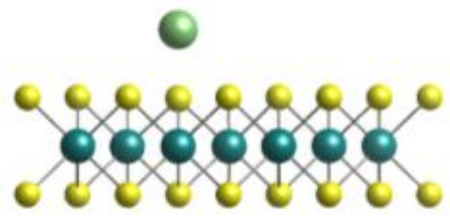 -0.96 | 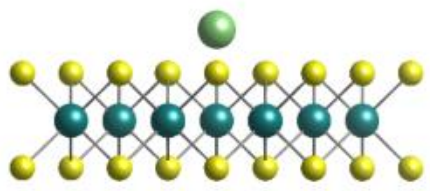 -1.69 |

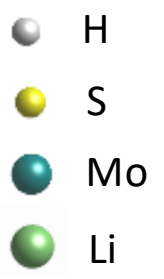

To explore the photophysics after chemical treatments, we conducted time-resolved PL (TRPL) and ultrafast pump-probe measurements. Normalized TRPL decays of H-TFSI-treated and Li-TFSI-treated $MoS_2$ samples at room temperature show noticeably different exciton decay dynamics (Fig. 3a). The TRPL curves are fitted by a three-exponential decay function with average lifetime ($\langle\tau\rangle$) ~320 ps, and ~150 ps for H-TSFI-treated and Li-TFSI-treated $MoS_2$, respectively. The TRPL decay of pristine $MoS_2$ is not presented as it is below the instrument response function (IRF) limit (~100 ps). At room temperature, the decay components can be attributed to a variety of sources.[37] The longer lifetime upon

H-TFSI treatment indicates a trap mediated exciton recombination process, which has been discussed in detail in previous studies and is supported by the following ultrafast pump-probe measurements.[38] In contrast, the shorter PL lifetime in Li-TFSI-treated $MoS_2$ indicates a greatly reduced role of exciton traps and is again consistent with the pump-probe results to follow.

The pump-probe spectra of pristine, H-TFSI-treated, and Li-TFSI-treated $MoS_2$ are depicted in Fig. S10, Fig. 3b and Fig. 3c, respectively. The exciton dynamics of pristine and H-TFSI-treated $MoS_2$ samples have been discussed in detail in our previous study.[39] The principle ground state bleach features correspond to the A and B excitons, at around 660 nm and 600 nm respectively.[40] The exciton lifetime is lengthened in the H-TFSI-treated sample due to the repopulation of the A exciton *via* thermal activation out of trap sites related to sulphur vacancies. These sub-gap trap sites (sulphur vacancies), which appear as a positive feature at 730 nm in the pump-probe (ΔT/T) spectra, have been previously detected in H-TFSI-treated $MoS_2$ and been shown to lead to trap limited emission lifetimes. In contrast, no sub-gap defect state emerges in the pump-probe spectra of the Li-TFSI-treated $MoS_2$ sample (no photo-induced features seen at 730 nm), indicating the lack of sub-gap trap sites which then leads to a shorter exciton lifetime. The pump-probe results agree well with TRPL data where Li-TFSI-treated $MoS_2$ sample presents a shorter lifetime due to a lack of exciton trapping and suggests that excitons in Li-TFSI-treated $MoS_2$ recombine more efficiently bypassing trap states or that the subgap state formed due to sulphur vacancies is passivated.

Exciton transport is an important criteria in many optoelectronic device and one that can be strongly affected by semiconductor properties such as doping and traps. Here, we directly monitor the spatial propagation of photogenerated excitons in H-TFSI-treated and Li-TFSI-treated $MoS_2$ monolayers on quartz substrates under ambient conditions with a confocal PL set up, as shown in Fig. 3b and 4c.[41–43] The Gaussian pump beam creates an Gaussian initial distribution population of excitons $n\,(x, 0)$ created by at position ($x_0$), which is given by[44]

$$n(x,0) = N \exp\left[-\frac{(x-x_0)^2}{2\sigma_0^2}\right] \tag{1}$$

With a variance of $\sigma_0^2$. In the following, the exciton density at any delay time ($t$) will be approximated with another Gaussian function:

$$n(x,t) = N \exp\left[-\frac{(x-x_0)^2}{2\sigma_t^2}\right] \tag{2}$$

With a variance of $\sigma_t^2$. The normalized PL intensity profile ($I_{PL}$) at each time snapshot ($t$) for H-TFSI-treated and Li-TFSI-treated monolayer $MoS_2$ are shown in Fig. 3d and 3e, respectively, together with the instrument response. For any considered time, the normalized PL profiles is well fitted with the Gaussian model. This allows us to extract the time evolution of the variance $\sigma_t^2$ for the two samples. At early time, (t<1 ns), $\sigma_t^2$ grows linearly with time, which is indicative of a diffusive motion of excitons.[45] At longer time, the value of $\sigma_t^2$ tends to saturate or even decrease for H-TFSI-treated $MoS_2$ sample. This indicates that the majority of propagating excitons has already decayed and that remaining ones are located around the point of creation ($x$ = 0). From the diffusive part of the curve, the exciton diffusion coefficient ($D$) is extracted from the slope of the fitting lines (Fig. 3f), using the diffusion equation:

$$D = \frac{\sigma_t^2 - \sigma_0^2}{2t} \tag{3}$$

The higher $D_{\text{Li-TFSI}}$ value of 0.22 cm$^2$ s$^{-1}$ in the Li-TFSI-treated $MoS_2$ sample compared to $D_{\text{H-TFSI}}$ value of 0.1 cm$^2$ s$^{-1}$ in H-TFSI-treated $MoS_2$ sample indicates that excitons in Li-TFSI-treated $MoS_2$ sample propagate more efficiently without trapping.

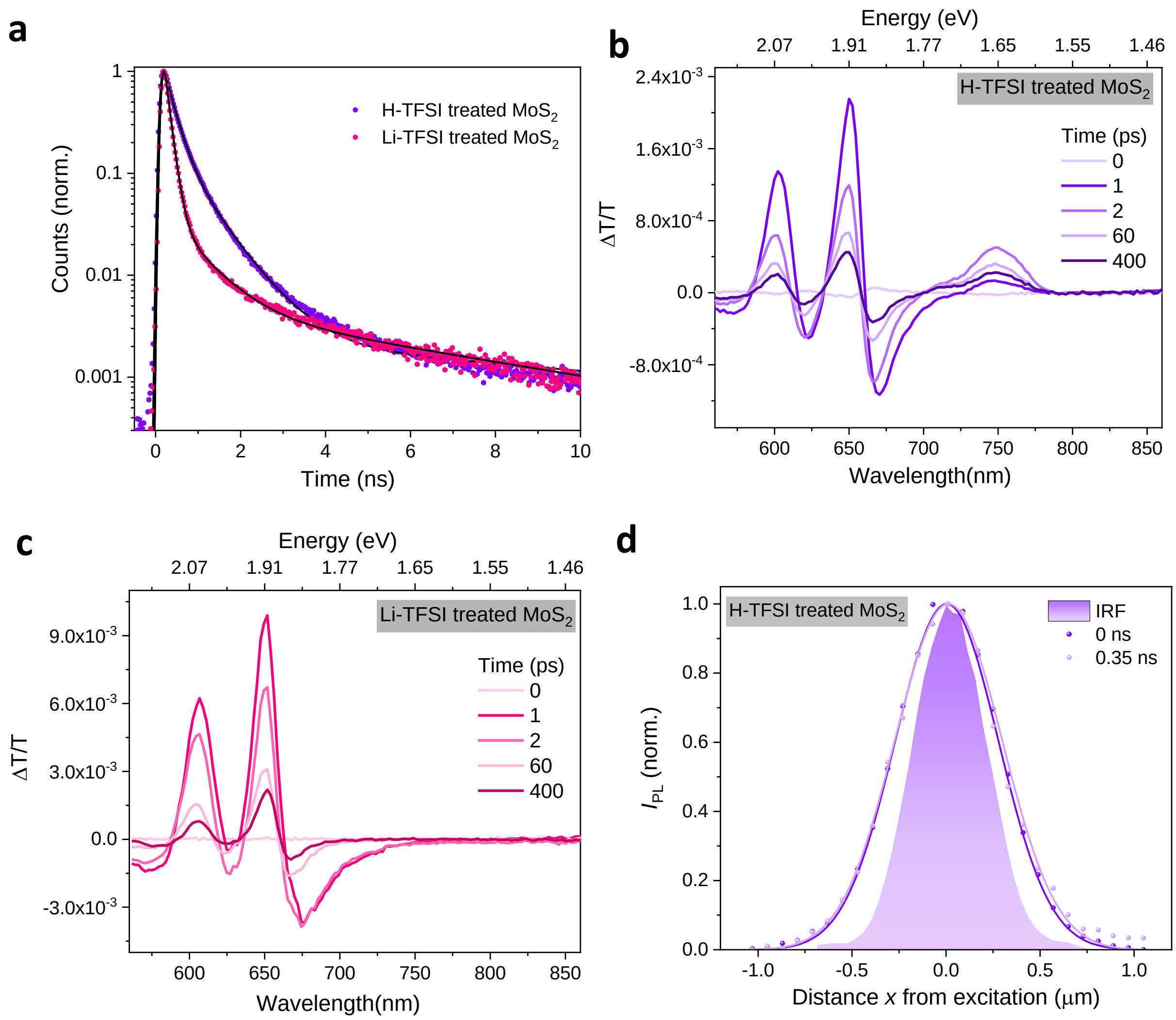

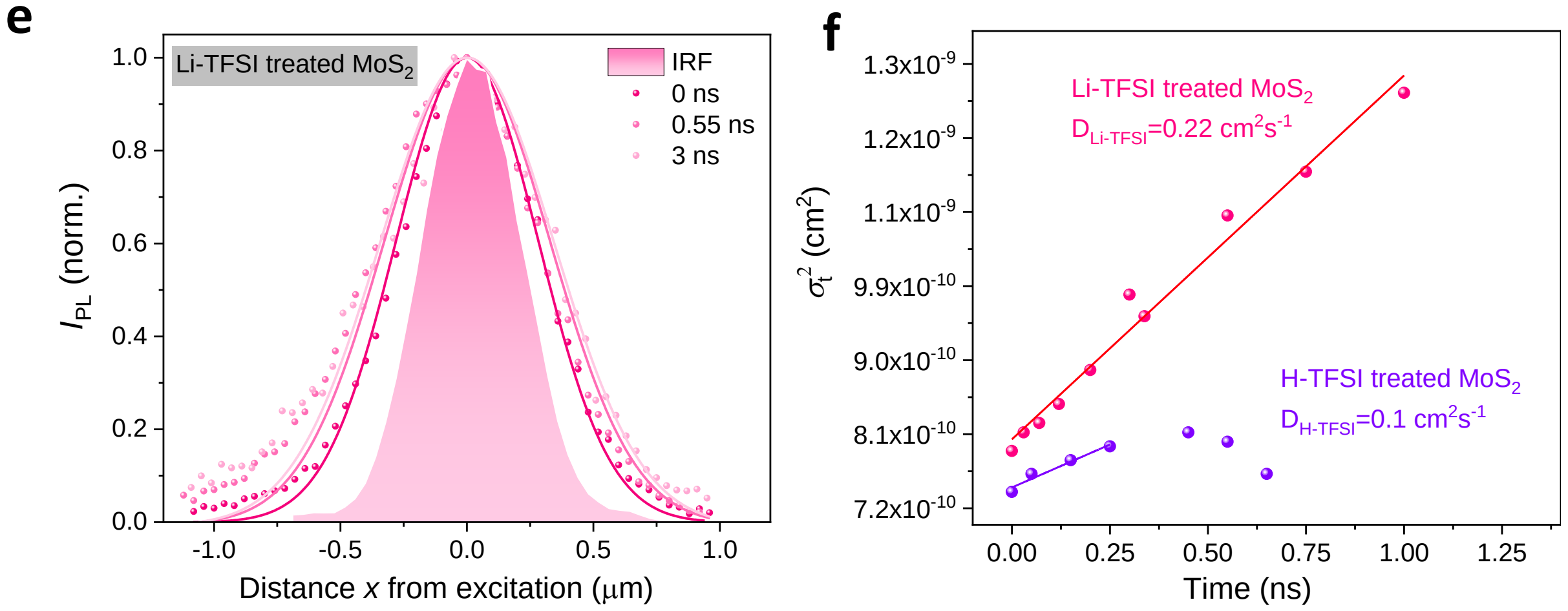

**Fig. 3 Time-resolved photoluminescence (TRPL), pump-probe spectra and photoluminescence propagation (diffusion) of treated $MoS_2$ monolayers. a** TRPL decay curves for H-TFSI-treated and Li-TFSI-treated monolayer $MoS_2$. **b** pump probe data of H-TFSI-treated $MoS_2$ where features related to traps can be seen at 730 nm. **c** pump probe data of Li-TFSI-treated $MoS_2$ which show no trap related features at 730 nm. **d** Spatial profile of the normalized PL intensity $I_{PL}$ for time snapshots $t$ = 0 and 0.35 ns for H-TFSI-treated monolayer $MoS_2$. **e** Spatial profile of the normalized PL intensity $I_{PL}$ for time snapshot $t$ =0, 0.55 and 3 ns for Li-TFSI-treated monolayer $MoS_2$. **f** Variance $\sigma_t^2$ as a function of time extracted from the Gaussian PL diffusion profiles of Li-TFSI-treated and H-TFSI-treated $MoS_2$ samples. Diffusion coefficient ($D$) is obtained from fits to the diffusion plots.

In order to comprehensively understand the treatment mechanism, we also compare the PL intensity enhancements of TMDSs treated with $Li^+$ and $Na^+$ salts different counter anions. Lithium triflate (Li-Tf) and sodium triflate (Na-Tf) were employed for comparison in this work since the Tf anion shows great similarity to the TFSI anion and dissociates freely in solution. Lithium acetate (Li-OAc) was also selected to further explore the effect of the counter anions on PL modulation of TMDSs. The scatter plots of emission peak position and peak PL counts from PL maps of treated monolayer $MoS_2$ and $WS_2$ are shown in Fig. S11. The PL of Na-Tf-treated and Li-OAc-treated $MoS_2$ showed no observable PL enhancement, hence this data is not presented. Representative PL spectra for Li-Tf-treated monolayers $MoS_2$ and $WS_2$ are shown in Fig. 4. Li-Tf treatment presents a clear PL enhancement for both $MoS_2$ and $WS_2$, whereas Li-OAc treatment only increases PL of $WS_2$ sample slightly. The effect of Li-OAc on $MoS_2$ PL enhancement is difficult to determine since the PL of both pristine and Li-OAc-treated $MoS_2$ were unmeasurable. However, the improvement factors for both Li-Tf and Li-OAc are quite small compared to Li-TFSI treatment. Moreover, there is clear trion emission contribution in the Li-Tf-treated $MoS_2$ at 664 nm. The results clearly suggest that counter anions play an important role in modulating the PL of TMDSs. The DFT simulations of Tf and TFSI anion adsorption at the sulphur vacancy sites of monolayer $MoS_2$ show that Tf anion tends to fill in the sulphur vacancy whereas there is no interaction between TFSI anions and the $MoS_2$ surface (Fig. S12).

Based on the results presented here, we speculate there are two reasons why TFSI based ionic salts work so well to enhance the PL of TMDSs. The presence of two strong electron-withdrawing groups ($-CF_3SO_2$) on the same nitrogen atom leads to a significantly lower surface charge density of the TFSI anion compared to the Tf anion.[46] In addition, the bulky side groups ($-CF_3SO_2$) lead to huge steric hindrance and make TFSI non-coordinating, while the Tf anion can coordinate to Mo or W at surfaces of TMDSs and behave as a n-doping reagent. For example, in the case of Na-Tf treatment of $WS_2$, the PL decreased due to the negligible effect of the $Na^+$ cation and negative effect of n-doping of the Tf anion. The weak effect of Li-OAc is, on the other hand, explained by the weak dissociation of ions. As illustrated in Fig. S13a, the $A_{1g}$ and 2LA Raman modes for Li-Tf-treated $MoS_2$ are blueshifted due to the p-doping effect, and an $A_{2u}$ mode emerges due to $Li^+$ adsorption. In contrast, an $A_{2u}$ mode does not appear in Na-Tf and Li-OAc-treated $MoS_2$ samples (Fig. S13b, c), suggesting that the superior PL enhancement effect of Li-TFSI treatment is due to stable adsorption of Li adatom, and low surface charge density as well as non-coordinating nature of TFSI counter anion. The TRPL and PL diffusion measurements were carried out on Li-Tf-treated $MoS_2$ samples to further uncover the role of counter anion play in chemical treatment, as depicted in Fig. S14. The normalized TRPL decay curve is fitted by a three-exponential decay function with $\langle\tau\rangle$ ~160 ps showing no evidence of surface trapping. The low $D_{Li\text{-}Tf}$ extracted of 0.12 $cm^2\ s^{-1}$ are, therefore, ascribed to collision of excitons with excess electrons during the diffusion process.[47]

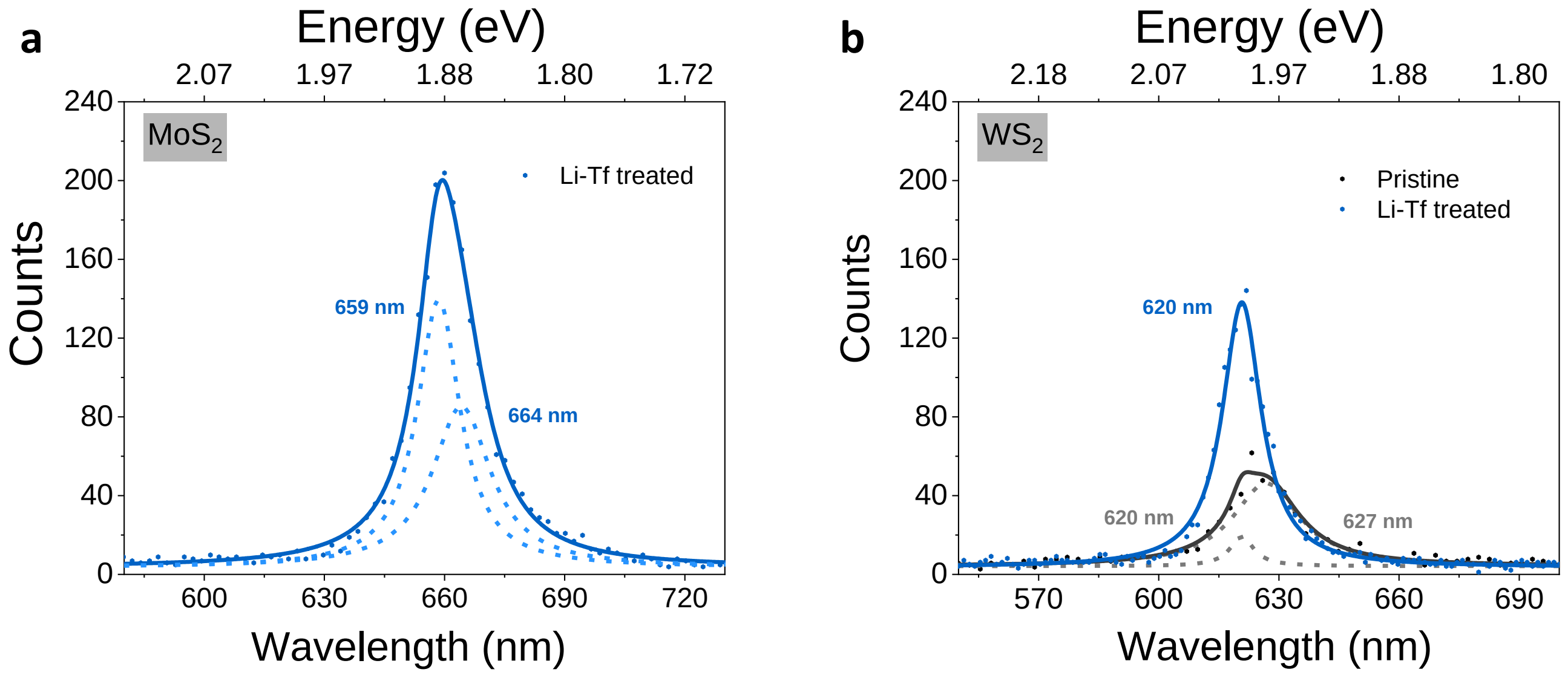

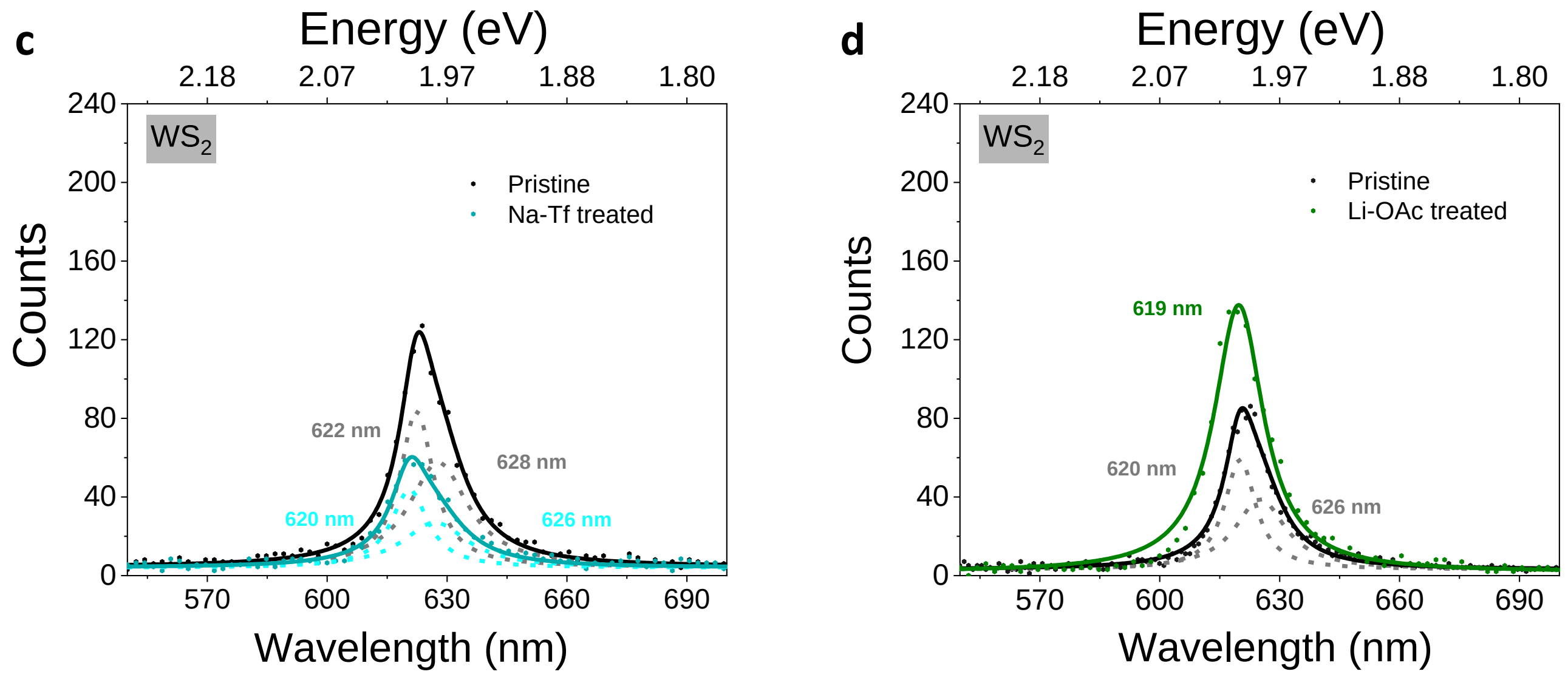

**Fig. 4 Photoluminescence (PL) spectra of $M_3$-Tf ($M_3$=Li and Na) and Li-OAc treated $MoS_2$ and $WS_2$ monolayers.** Representative PL spectra for **a** Li-Tf-treated monolayer $MoS_2$, **b** pristine and Li-Tf-treated monolayer $WS_2$, **c** pristine and Na-Tf-treated monolayer $WS_2$, and **d** pristine and Li-OAc-treated monolayer $WS_2$. The decomposed Lorentzian peak fitting is presented in dashed line and cumulative Lorentzian peak fittings are presented in solid line.

## Conclusions

We have systematically investigated surface chemical treatments that enhance the PL yield of TMDSs by comparing a series of ionic chemicals and small molecule p-dopants, and studying their effect via a range of steady-state and time-resolved spectroscopy and microscopy techniques combined with DFT simulations. Our results provide a detailed mechanistic picture for how these chemical treatments work and allow us to set up selection rules for ionic chemicals to improve PL of TMDSs, where cations and counter anions both play important roles during chemical treatments. The cation must be stably adsorbed on the surface of TMDSs rather than underdo electron transfer, allowing for suppression of trion formation, thereby improving PL yield. The counter anion should be non-coordinating with strong electron-withdrawing groups. The strongest enhancement is observed for Li-TFSI, which gives a PL enhancement twice that of the widely discussed “super acid” H-TFSI. More importantly, Li-TFSI is stable and functions in benign solvents, which possesses the potential to be employed directly during device fabrication of TMDSs. Overall, we demonstrate a simple and effective route to enhance PL of TMDSs which opens a route to building high performance chemically treated optoelectronic devices.

## Methods

### Material

Bulk $MoS_2$ and $WS_2$ crystals were purchased from 2D Semiconductors. The monolayer $MoS_2$ and $WS_2$ were prepared according to reported gold-mediated exfoliation method to ensure relatively large monolayers.[48] In this study, all experiments were carried out on monolayers. All chemicals for the surface treatments were purchased from Sigma-Aldrich and used as received.

### Chemical treatments

The chemical treatments with H-TFSI (0.02 M in 1, 2-dichloroethane), F4TCNQ (0.02 M in dichloromethane) and Magic Blue (0.02 M in dichloromethane) are carried out inside a nitrogen glovebox, and other treatments are carried out in ambient atmosphere. Methanol is used as solvent for all ionic salts for comparison. The chemical treatments were achieved by immersing the samples into concentrated solutions of the investigated chemicals (0.02 M) for 40 mins.

### Characterization

The microscope steady-state PL measurement was carried out using a WITec alpha 300 s setup and has been described previously.[49] Importantly, a 405 nm continuous wave laser (Coherent CUBE) was used as the excitation source. A long pass filter with a cutoff wavelength of 450 mm was fitted before signal collection to block excitation scatter. The light was coupled with an optical fiber to the microscope and focused using a 20× Olympus lens. Samples were placed on an X-Y piezo stage of the microscope. The PL signal was collected in refection mode with the same 20× objective and detected using a Princeton Instruments SP-2300i spectrometer fitted with an Andor iDus 401 CCD detector. The PL maps were measured at 405 nm excitation with a fluence of 15 W $cm^{-2}$. The Raman measurements were carried out using Renishaw inVia Raman confocal microscope with a 532 nm excitation source. The XPS measurements were performed using a Thermo Escalab 250Xi system and monochromated aluminium $K_\alpha$ x-ray source. The software package “Thermo Avantage” (Thermo Fisher Scientific Inc., Waltham, USA) was used for data analysis.

The ultrafast pump-probe setup has been described previously.[50] A Light Conversion PHAROS laser system with 400 μJ per pulse at 1030 nm with a repetition rate of 38 kHz is split in two, one part is used to generate the continuum probe light and the second part is used in an Collinear Optical Parametric Amplifier (Orpheus, Light Conversion) to generate the pump source at the desired wavelength. The probe pulse is delayed up to 2 ns with a mechanical delay-stage (Newport). A mechanical chopper (Thorlabs) is used to create an on-off pump-probe pulse series. A silicon line scan camera (JAI SW-2000M-CL-80) fitted onto a visible spectrograph (Andor Solis, Shamrock) is used to record the transmitted probe light. The time-resolved photoluminescence (TRPL) microscopy measurements were performed using 405 nm pulsed laser (PDL 828-S "SEPIA II", PicoQuant) excitation *via* 100× objective in a PicoQuant Microtime 200 confocal setup. The emission signal was separated from the excitation light using a dichroic mirror (Z405RDC, Chroma). The TRPL was measured at 15 μJ $cm^{-2}$ and data was averaged from 100 $\mu m^2$ monolayer flakes. PL signals were collected in transmission mode and instrument response functions (IRF) were measured with blank quartz substrates. For the diffusion measurements, the emission path was raster scanned while the excitation was decoupled and fixed at the center of the sampler ($x = 0$). The PL was then focused onto a Hybrid PMT detector (Picoquant) for single-photon counting (time resolution of 60 ps) through a pinhole (50 μm), with an additional 410-nm longpass filter. Repetition rates of 27 MHz were used for the maps and the diffusion profiles. The lateral spatial resolution is ~550 nm. An incident power of 60 nW was used, corresponding to a fluence of 700 nJ $cm^{-2}$.

**Supporting Information**

Supporting Information is available online with additional experimental and calculation details as well as additional data for optical and photophysical characterizations of different chemical treatments on $MoS_2$ and $WS_2$.

Data available in University of Cambridge data repository at: link to be added during proof.

No custom computer code is used in this work.

**Acknowledgement**

This project has received funding from the European Research Council (ERC) under the European Union's Horizon 2020 research and innovation program (Grant Agreement No. 758826 & 756962). Z.L. acknowledges funding from the Swedish research council, Vetenskapsrådet 2018-06610. S.D.S. acknowledges funding from the Royal Society and Tata Group (UF150033). G.D. acknowledges the Royal Society for funding through a Newton International Fellowship. We acknowledge financial support from the EPSRC and the Winton Programme for the Physics of Sustainability.

**Reference**

1. Jariwala, D., Sangwan, V. K., Lauhon, L. J., Marks, T. J. & Hersam, M. C. Emerging device applications for semiconducting two-dimensional transition metal dichalcogenides. *ACS Nano* **8**, 1102–1120 (2014).
2. Mak, K. F. & Shan, J. Photonics and optoelectronics of 2D semiconductor transition metal

dichalcogenides. *Nat. Photonics* **10**, 216–226 (2016).
3. Tan, C. *et al.* Recent Advances in Ultrathin Two-Dimensional Nanomaterials. *Chem. Rev.* **117**, 6225–6331 (2017).
4. Chhowalla, M. *et al.* The chemistry of two-dimensional layered transition metal dichalcogenide nanosheets. *Nat. Chem.* **5**, 263–275 (2013).
5. Su, L. *et al.* Inorganic 2D Luminescent Materials: Structure, Luminescence Modulation, and Applications. *Adv. Opt. Mater.* **1900978**, 1900978 (2019).
6. Hong, J. *et al.* Exploring atomic defects in molybdenum disulphide monolayers. *Nat. Commun.* **6**, 1–8 (2015).
7. Qiu, H. *et al.* Hopping transport through defect-induced localized states in molybdenum disulphide. *Nat. Commun.* **4**, 3–8 (2013).
8. Mak, K. F. *et al.* Tightly bound trions in monolayer MoS 2. *Nat. Mater.* **12**, 207–211 (2013).
9. Lien, D. H. *et al.* Electrical suppression of all nonradiative recombination pathways in monolayer semiconductors. *Science (80-. ).* **364**, 468–471 (2019).
10. Zheng, W. *et al.* Light Emission Properties of 2D Transition Metal Dichalcogenides: Fundamentals and Applications. *Adv. Opt. Mater.* **6**, 1–29 (2018).
11. Zhang, S. *et al.* Controllable, Wide-Ranging n-Doping and p-Doping of Monolayer Group 6 Transition-Metal Disulfides and Diselenides. *Adv. Mater.* **30**, (2018).
12. Radisavljevic, B., Radenovic, A., Brivio, J., Giacometti, V. & Kis, A. Single-layer MoS2 transistors. *Nat. Nanotechnol.* **6**, 147–150 (2011).
13. Bertolazzi, S., Gobbi, M., Zhao, Y., Backes, C. & Samorì, P. Molecular chemistry approaches for tuning the properties of two-dimensional transition metal dichalcogenides. *Chem. Soc. Rev.* **47**, 6845–6888 (2018).
14. Voiry, D. *et al.* Covalent functionalization of monolayered transition metal dichalcogenides by phase engineering. *Nat. Chem.* **7**, 45–49 (2015).
15. Tanoh, A. O. A. *et al.* Enhancing Photoluminescence and Mobilities in WS2 Monolayers with Oleic Acid Ligands. *Nano Lett.* **19**, (2019).
16. Bonilla, R. S., Hoex, B., Hamer, P. & Wilshaw, P. R. Dielectric surface passivation for silicon solar cells: A review. *Phys. Status Solidi Appl. Mater. Sci.* **214**, (2017).
17. Aberle, A. G. Surface Passivation of Crystalline silicon solar cells: A Review. *Prog. Photovoltaics Res. Appl.* **8**, 473–487 (2000).
18. Peimyoo, N. *et al.* Chemically driven tunable light emission of charged and neutral excitons in monolayer WS 2. *ACS Nano* **8**, 11320–11329 (2014).
19. Mouri, S., Miyauchi, Y. & Matsuda, K. Tunable photoluminescence of monolayer MoS2 via

chemical doping. *Nano Lett.* **13**, 5944–5948 (2013).

20. Amani, M. *et al.* Near-unity photoluminescence quantum yield in MoS2. *Science (80-. ).* **350**, 1065–1068 (2015).
21. Yu, Z. *et al.* Towards intrinsic charge transport in monolayer molybdenum disulfide by defect and interface engineering. *Nat. Commun.* **5**, 1–7 (2014).
22. Lu, H., Kummel, A. & Robertson, J. Passivating the sulfur vacancy in monolayer MoS2. *APL Mater.* **6**, (2018).
23. Schwermann, C. *et al.* Incorporation of oxygen atoms as a mechanism for photoluminescence enhancement of chemically treated MoS2. *Phys. Chem. Chem. Phys.* **20**, 16918–16923 (2018).
24. Kang, N., Paudel, H. P., Leuenberger, M. N., Tetard, L. & Khondaker, S. I. Photoluminescence quenching in single-layer MoS2 via oxygen plasma treatment. *J. Phys. Chem. C* **118**, 21258–21263 (2014).
25. Christopher, J. W., Goldberg, B. B. & Swan, A. K. Long tailed trions in monolayer MoS2: Temperature dependent asymmetry and resulting red-shift of trion photoluminescence spectra. *Sci. Rep.* **7**, 1–8 (2017).
26. Cadiz, F. *et al.* Well separated trion and neutral excitons on superacid treated MoS2 monolayers. *Appl. Phys. Lett.* **108**, (2016).
27. Lee, J.-U., Kim, M. & Cheong, H. Raman Spectroscopic Studies on Two-Dimensional Materials. *Appl. Microsc.* **45**, 126–130 (2015).
28. Guo, H. *et al.* Resonant Raman spectroscopy study of swift heavy ion irradiated MoS2. *Nucl. Instruments Methods Phys. Res. Sect. B Beam Interact. with Mater. Atoms* **381**, 1–5 (2016).
29. Qian, Q., Zhang, Z. & Chen, K. J. In Situ Resonant Raman Spectroscopy to Monitor the Surface Functionalization of MoS2and WSe2for High-k Integration: A First-Principles Study. *Langmuir* **34**, 2882–2889 (2018).
30. Dhakal, K. P. *et al.* Heterogeneous modulation of exciton emission in triangular WS2 monolayers by chemical treatment. *J. Mater. Chem. C* **5**, 6820–6827 (2017).
31. Tarasov, A. *et al.* Controlled doping of large-area trilayer MoS2with molecular reductants and oxidants. *Adv. Mater.* **27**, 1175–1181 (2015).
32. Diao, Y., Xie, K., Xiong, S. & Hong, X. Insights into Li-S battery cathode capacity fading mechanisms: Irreversible oxidation of active mass during cycling. *J. Electrochem. Soc.* **159**, 1816–1821 (2012).
33. Eda, G. *et al.* Photoluminescence from chemically exfoliated MoS 2. *Nano Lett.* **11**, 5111–5116 (2011).
34. Manuja, M., Sarath Krishnan, V. & Jose, G. Molybdenum Disulphide Nanoparticles Synthesis

Using a Low Temperature Hydrothermal Method and Characterization. *IOP Conf. Ser. Mater. Sci. Eng.* **360**, (2018).

35. Hussain, S. *et al.* Large-area, continuous and high electrical performances of bilayer to few layers MoS2 fabricated by RF sputtering via post-deposition annealing method. *Sci. Rep.* **6**, 1–13 (2016).
36. Barik, G. & Pal, S. Defect Induced Performance Enhancement of Monolayer MoS 2 for Li- and Na-Ion Batteries . *J. Phys. Chem. C* **123**, 21852–21865 (2019).
37. McCreary, K. M., Currie, M., Hanbicki, A. T., Chuang, H. J. & Jonker, B. T. Understanding Variations in Circularly Polarized Photoluminescence in Monolayer Transition Metal Dichalcogenides. *ACS Nano* **11**, 7988–7994 (2017).
38. Goodman, A. J., Willard, A. P. & Tisdale, W. A. Exciton trapping is responsible for the long apparent lifetime in acid-treated MoS2. *Phys. Rev. B* **96**, 1–6 (2017).
39. Bretscher, H. M., Li, Z., Xiao, J., Qiu, D. Y. & Refaely-abramson, S. The bright side of defects in MoS 2 and WS 2 and a generalizable chemical treatment protocol for defect passivation. *Arxiv:2002.03956* (2020).
40. Dal Conte, S., Trovatello, C., Gadermaier, C. & Cerullo, G. Ultrafast Photophysics of 2D Semiconductors and Related Heterostructures. *Trends Chem.* 1–15 (2019) doi:10.1016/j.trechm.2019.07.007.
41. Stavrakas, C. *et al.* Visualizing buried local carrier diffusion in halide perovskite crystals via two-photon microscopy. *ACS Energy Lett.* **5**, 117–123 (2020).
42. Kulig, M. *et al.* Exciton Diffusion and Halo Effects in Monolayer Semiconductors. *Phys. Rev. Lett.* **120**, 207401 (2018).
43. Cadiz, F. *et al.* Exciton diffusion in WSe2 monolayers embedded in a van der Waals heterostructure. *Appl. Phys. Lett.* **112**, (2018).
44. Yuan, L., Wang, T., Zhu, T., Zhou, M. & Huang, L. Exciton Dynamics, Transport, and Annihilation in Atomically Thin Two-Dimensional Semiconductors. *J. Phys. Chem. Lett.* **8**, 3371–3379 (2017).
45. Akselrod, G. M. *et al.* Subdiffusive exciton transport in quantum dot solids. *Nano Lett.* **14**, 3556–3562 (2014).
46. Hekselman *et al.* Effect of calix[6]pyrrole anion receptor addition on properties of PEO-based solid polymer electrolytes doped with LiTf and LiTfSI salts. *Electrochim. Acta* **55**, 1298–1307 (2010).
47. Kato, T. & Kaneko, T. Transport dynamics of neutral excitons and trions in monolayer WS2. *ACS Nano* **10**, 9687–9694 (2016).

48. Desai, S. B. *et al.* Gold-Mediated Exfoliation of Ultralarge Optoelectronically-Perfect Monolayers. *Adv. Mater.* **28**, 4053–4058 (2016).

49. Tainter, G. D. *et al.* Long-Range Charge Extraction in Back-Contact Perovskite Architectures via Suppressed Recombination. *Joule* **3**, 1301–1313 (2019).

50. Allardice, J. R. *et al.* Engineering Molecular Ligand Shells on Quantum Dots for Quantitative Harvesting of Triplet Excitons Generated by Singlet Fission. *J. Am. Chem. Soc.* **141**, 12907–12915 (2019).

# Supporting Information

**Mechanistic Insight to the Chemical Treatments of Monolayer Transition Metal Disulphides for Photoluminescence Enhancement**

*Zhaojun Li, Hope Bretscher, Yunwei Zhang, Géraud Delport, James Xiao, Alpha Lee, Samuel D. Stranks, and Akshay Rao**

# Contents

# 1. Experimental details

Si-$SiO_2$ substrates with 90 nm oxide layer were used for steady-state photoluminescence (PL), Raman spectroscopy and X-ray photoemission spectroscopy (XPS). Quartz substrates were used for time-resolved photoluminescence (TRPL), ultrafast pump-probe measurement, and PL diffusion measurements. The samples were encapsulated for ultrafast pump-probe measurements, and other measurements are carried out on samples without encapsulation.

# 2. Calculation details

First-principle calculations of formation energies were carried out based on the density functional theory (DFT) with the Perdew-Burke-Ernerhof (PBE) exchange-correlation functional as implemented in the VASP code. The all-electron projector-augmented wave (PAW) method was adopted, where $4d^5 5s^1$, $5d^4 6s^2$, $3s^2 3p^4$, $1s^1$, $2s^1$, and $3s^1$ are treated as valence electrons for Mo, W, S, H, Li and Na atoms, respectively. The plane-wave energy cutoff is set to 600 eV. A Monkhorst-Pack Brillouin zone sampling grid with a resolution of $2\pi \times 0.03$ Å$^{-1}$ is adopted to ensure that all the enthalpy calculations are well converged with an error less than 1 meV/atom. Structural relaxations were performed with forces converged to less than 0.01 eV Å$^{-1}$. The 3×3 hexagonal supercell of monolayer $MoS_2$ and $WS_2$ were utilized to display various available adsorption sites in $MoS_2$/ $WS_2$ for $M_1$ (H, Li, Na) adsorption. A vacuum spacing of 20 Å was provided along a perpendicular direction to the plane of $MoS_2$/ $WS_2$ between two adjacent periodic layers in order to avoid any spurious interactions. Detailed structure information is listed below.

Taking $MoS_2$ as the example, the stability of various adsorption sites is calculated from their formation energy, which is defined as:

$$E = E_{M1+MoS_2} - E_{MoS_2} - E_{M1} \qquad (S1)$$

where $E_{M1+MoS_2}$ is the total energy of $M_1$ (H, Li and Na) adsorbed $MoS_2$, $E_{MoS_2}$ is the energy of $MoS_2$ before the adsorption and $E_{M1}$ is the energy of an isolated M1 atom. According to the definition, the structure with a more negative formation energy is more stable.

# 3. Additional PL data for chemical treated $MoS_2$ and $WS_2$

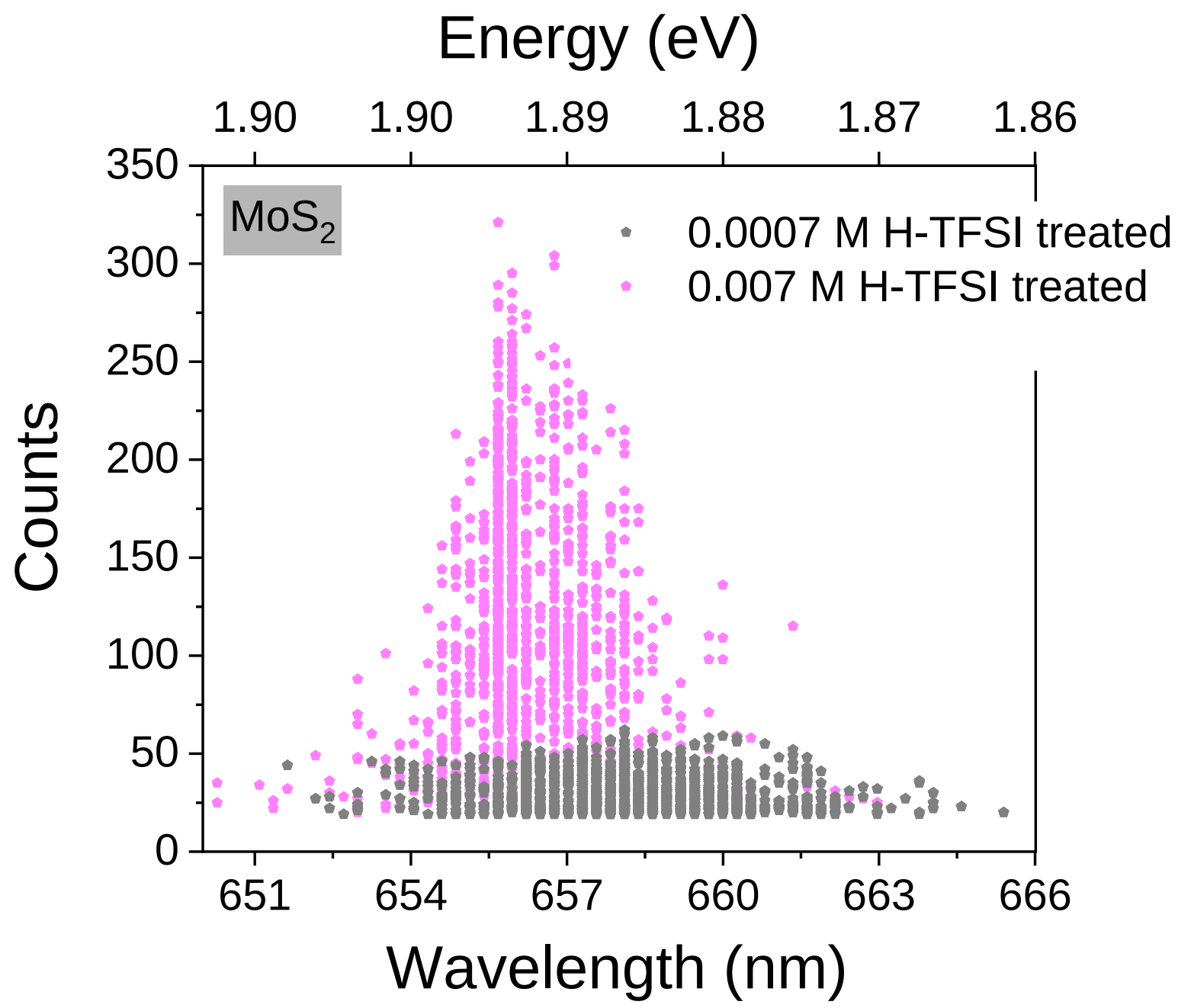

**Fig. S1** PL enhancement scatter plots of spectral position of the peak emission and peak H-TFSI-treated monolayer $MoS_2$ PL counts extracted from PL maps of $MoS_2$ monolayer on Si-$SiO_2$ (90 nm) after surface treatment with different concentrations of H-TFSI in 1, 2-dichloroethane.

The PL of pristine $MoS_2$ is usually undetectable due to the low PL intensity, and the corresponding statistic scatter plots of pristine $MoS_2$ are not presented. As the PL of pristine $WS_2$ is detectable, the corresponding pristine monolayer scatter plots are also shown in Fig. S2. For chemically treated $WS_2$ samples, we performed PL mapping on the same monolayer as on the pristine sample to obtain more direct observation of the PL enhancing strength of different chemical treatments. The PL of pristine $WS_2$ is inhomogeneous, and the position of PL maxima varies between 615 nm and 630 nm, which may be due to randomly-distributed disorder potentials, trions, dielectric disorder as well as interactions with optical phonons.[1–3] Upon chemical treatments, the PL of both $MoS_2$ and $WS_2$ increase and blue shift statistically, indicating an reduction of trions in both materials. This trend agrees well with previous observations reported by other groups.[4–6]

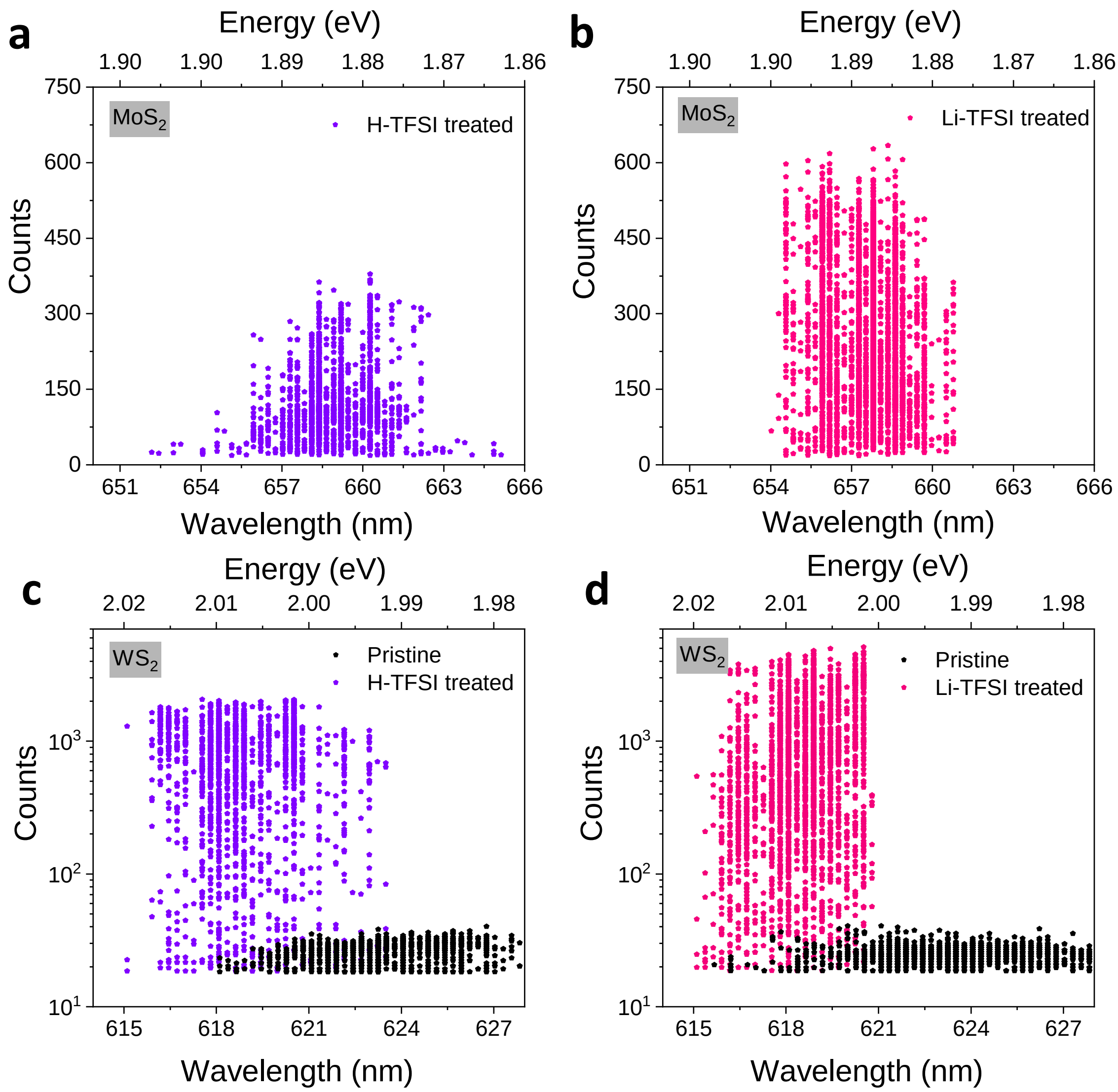

**Fig. S2** Photoluminescence scatter plots showing **a** peak H-TFSI-treated monolayer $MoS_2$ PL counts, **b** peak Li-TFSI-treated monolayer $MoS_2$ PL counts, **c** peak Na-TFSI-treated monolayer $MoS_2$ PL counts, **d** peak H-TFSI-treated and corresponding pristine monolayer $WS_2$ PL counts, **e** peak Li-TFSI-treated and corresponding pristine monolayer $WS_2$ PL counts, and **f** peak Na-TFSI-treated and corresponding pristine monolayer $WS_2$ PL counts. Data derived from raw spectra from PL maps.

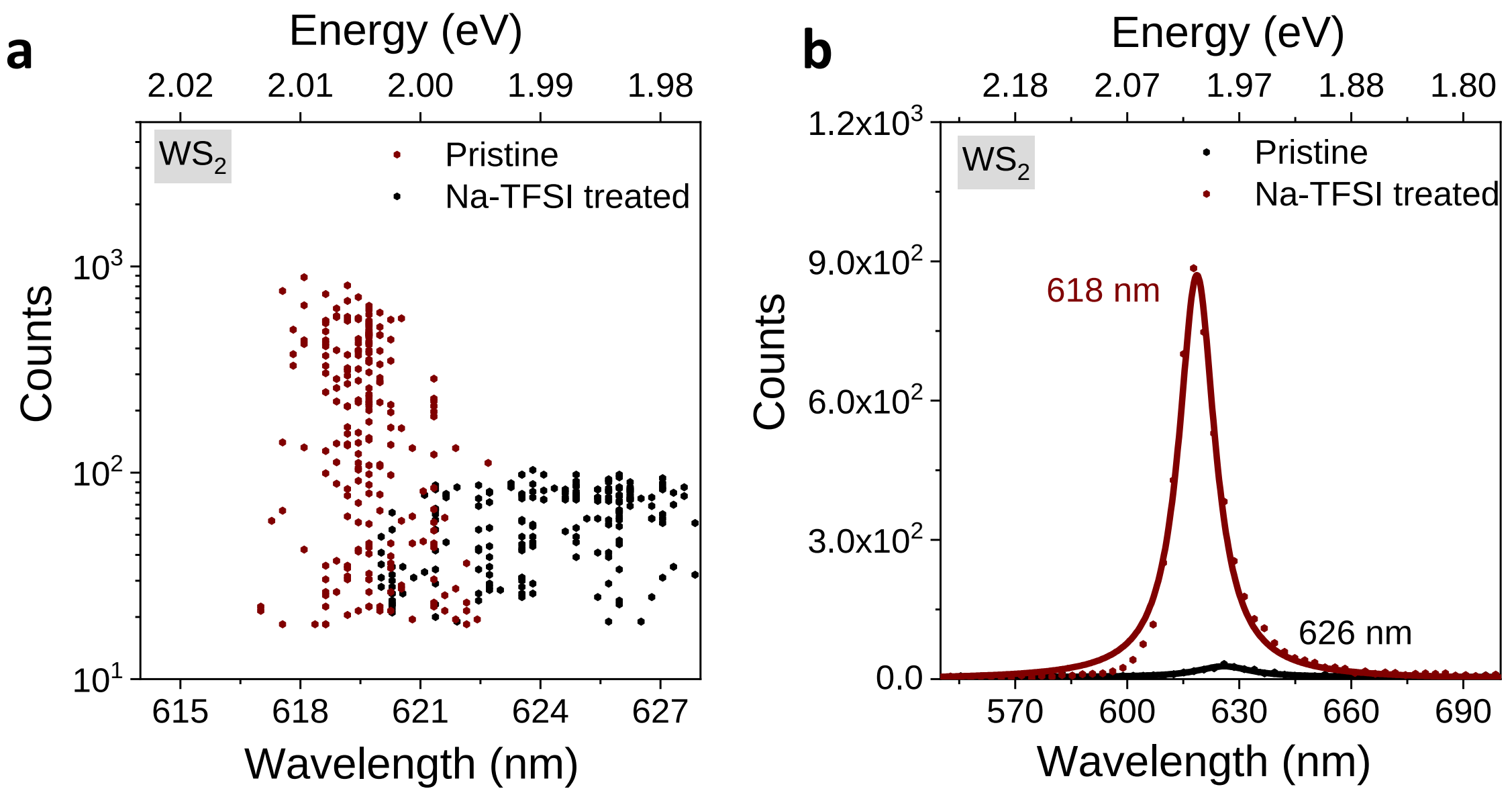

**Fig. S3 a** PL enhancement scatter plots showing peak Na-TFSI-treated monolayer $WS_2$ PL counts. **b** Maximum PL spectra for pristine and Na-TFSI-treated monolayer $WS_2$.

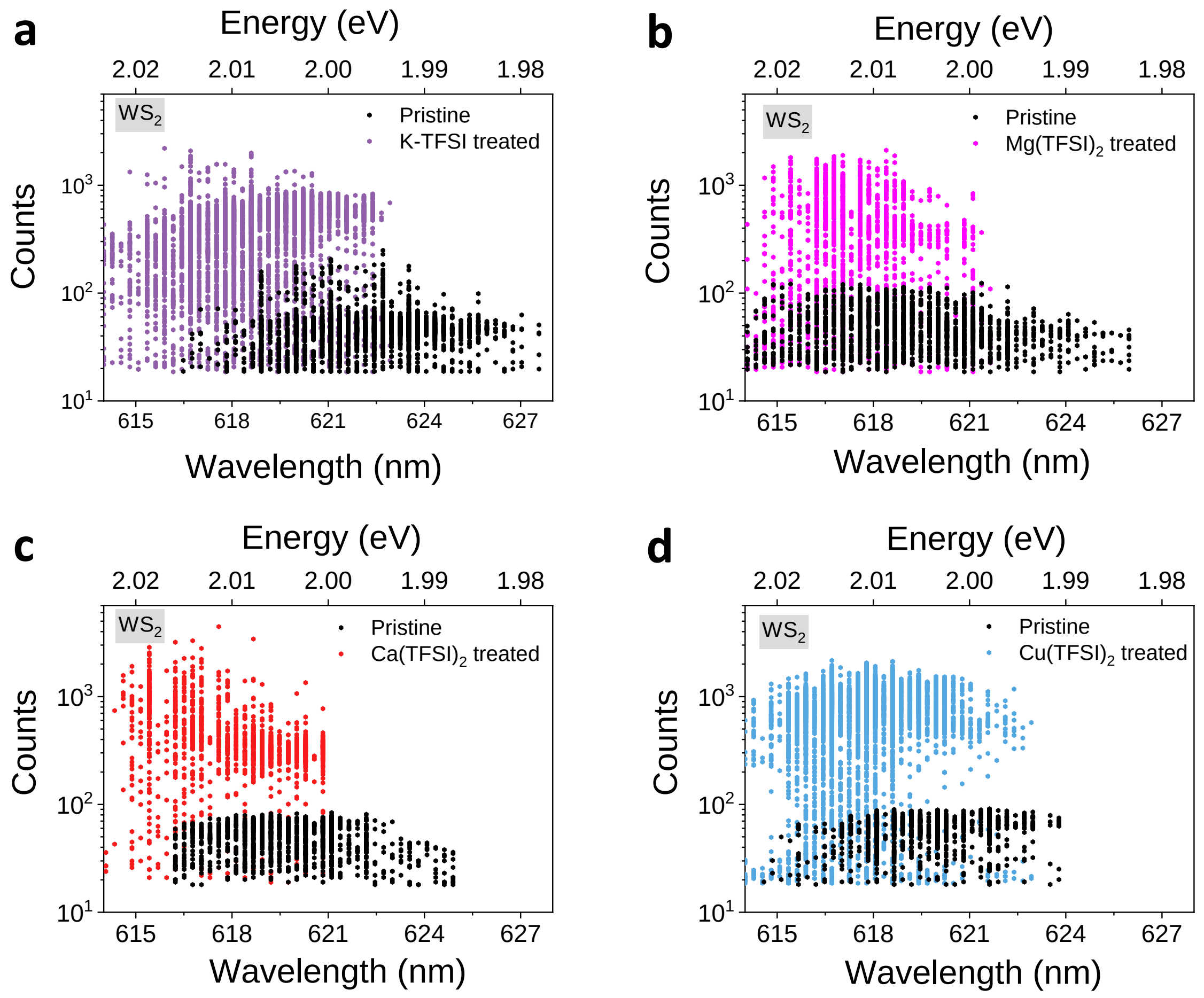

**Fig. S4 a** Chemical structures of K-TFSI and $M_2(TFSI)_2$ (M2 = Mg, Ca and Cu). PL scatter plots showing peak counts of **b** pristine and K-TFSI-treated $WS_2$, **c** pristine and $Mg(TFSI)_2$-treated $WS_2$, **d** pristine and $Ca(TFSI)_2$-treated $WS_2$, and **e** pristine and $Cu(TFSI)_2$-treated $WS_2$.

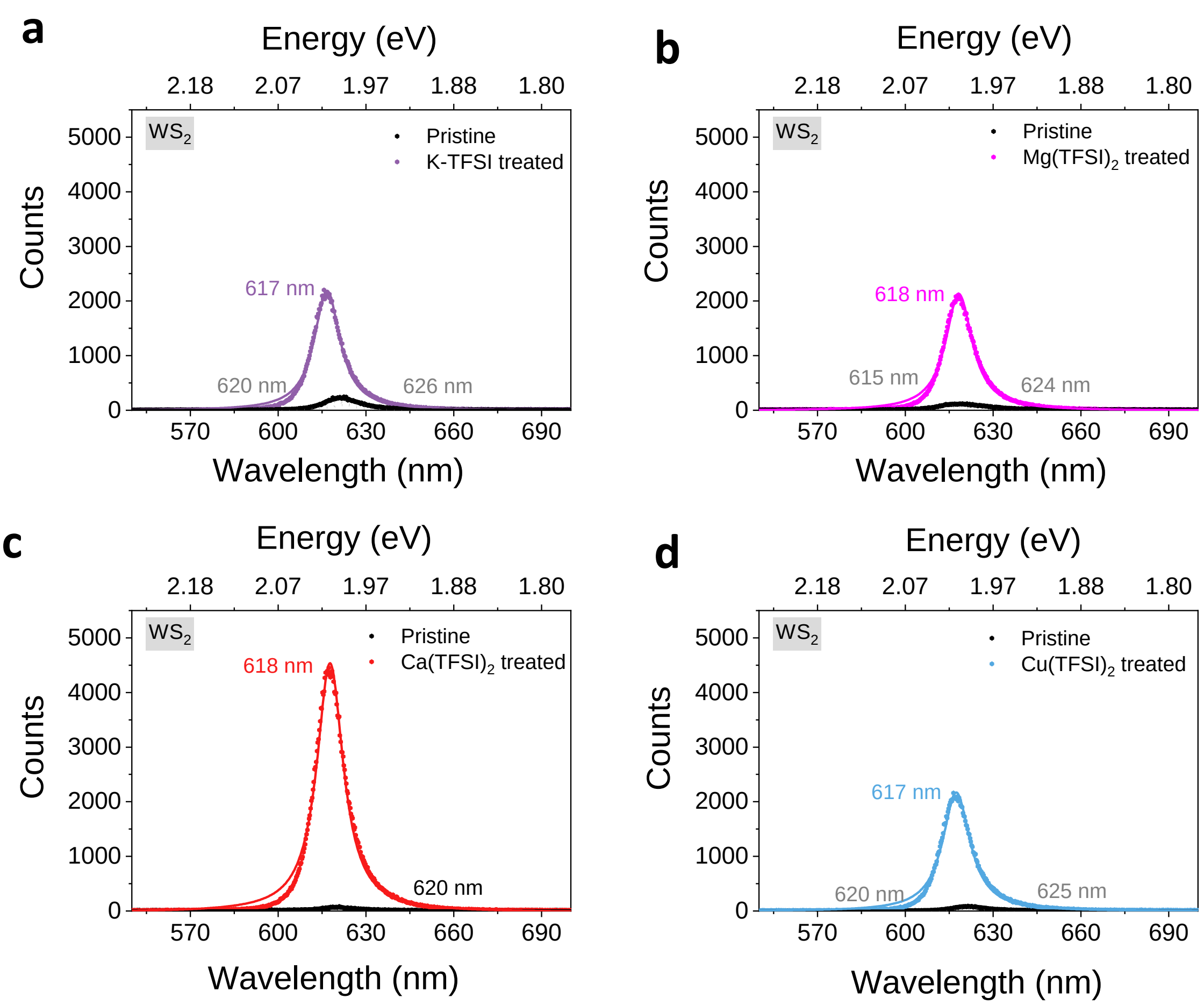

**Fig. S5 a** Maximum PL spectrum for pristine and K-TFSI-treated monolayer $WS_2$. **b** Maximum PL spectrum for pristine and $Mg(TFSI)_2$-treated monolayer $WS_2$. **c** Maximum PL spectrum for pristine and $Ca(TFSI)_2$-treated monolayer $WS_2$. **d** Maximum PL spectrum for pristine and $Cu(TFSI)_2$-treated monolayer $WS_2$. The decomposed Lorentzian peak fitting is presented in dash line and the cumulative peak fitting is presented in solid line.

To further evaluate if the cationic radii of TFSI salts play a role during the chemical treatments and if other TFSI salts can also enhance PL of TMDSs, we also investigated the effect of other five TFSI based ionic salts on the PL enhancement of $WS_2$. $Mg(TFSI)_2$ and $Cu(TFSI)_2$ show smaller cationic radii compared to Li-TFSI, while Na-TFSI, K-TFSI and $Ca(TFSI)_2$ show larger cationic radii compared to Li-TFSI.[7] As shown in Fig. S2-S5, these ionic salts all have positive effect on the PL of $WS_2$ and the treatments cause blueshift of PL spectra of $WS_2$.

Interestingly, the PL enhancement of $WS_2$ are similar by KTFSI, $Mg(TFSI)_2$ and $Ca(TFSI)_2$ treatments, although their cationic radii are quite different. Thus, no concrete relationship between cationic radii and PL tuning strength can be drawn.

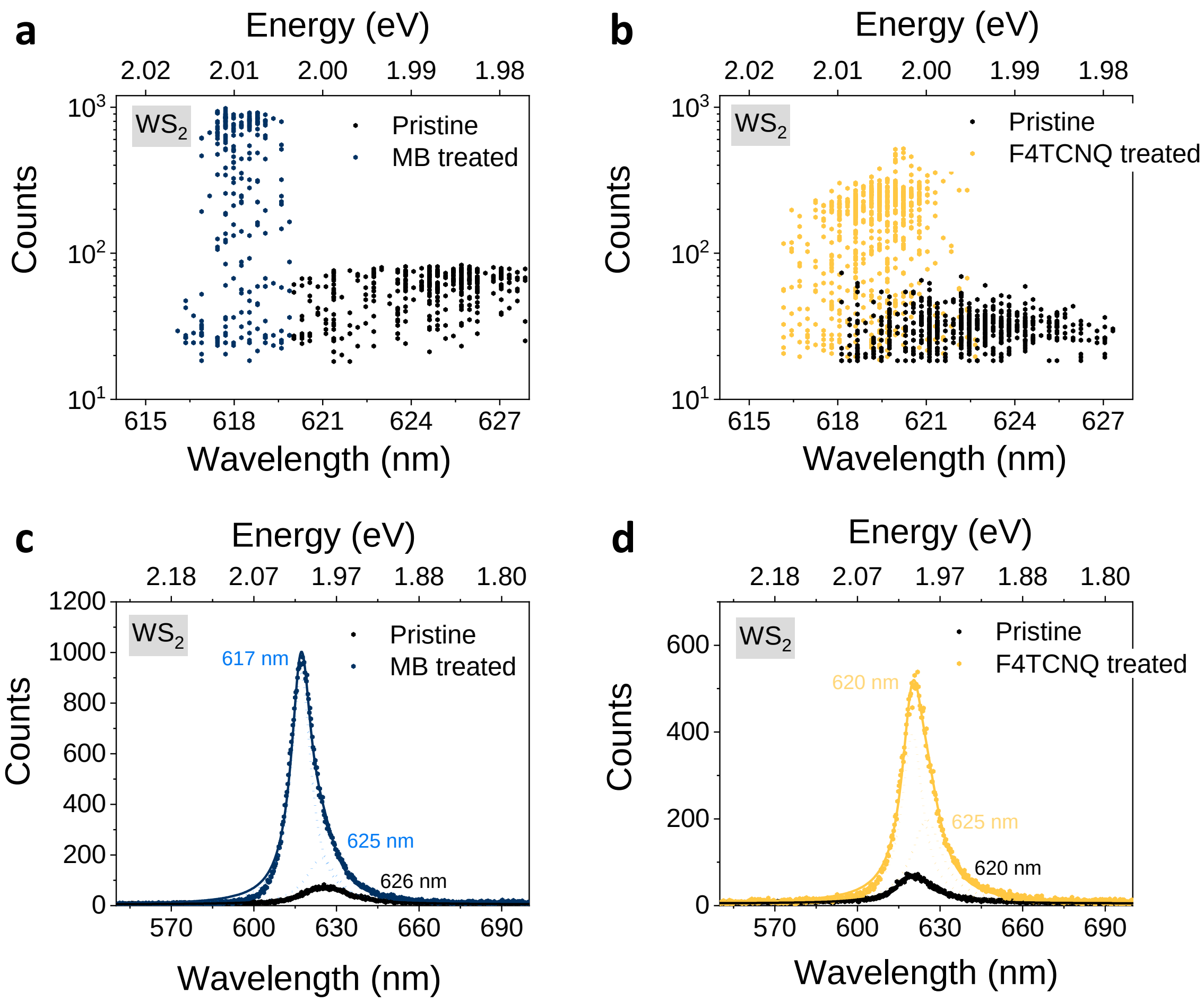

**Fig. S6 a** PL scatter plots showing peak pristine and MB-treated monolayer $WS_2$ counts. **b** PL scatter plots showing peak pristine and F4TCNQ-treated monolayer $WS_2$ PL counts. **c** Maximum PL spectrum for MB-treated monolayer $WS_2$. **d** Maximum PL spectrum for F4TCNQ-treated monolayer $WS_2$. The decomposed Lorentzian peak fitting of MB and F4TCNQ-treated $WS_2$ is presented in dash line and the cumulative peak fitting is presented in solid line.

The PL mappings of MB and F4TCNQ-treated $WS_2$ were performed on the same monolayers as on the pristine samples to obtain more direct observation of the PL enhancing strength of the chemical treatments. As shown in Fig. S5, both MB and F4TCNQ increased the PL of $WS_2$ slightly and blueshifted the PL spectra of $WS_2$. However, the enhancements are much weaker compared to H-TFSI and Li-TFSI treatments, and there are still clear trion contribution form the emission of MB and F4TCNQ-treated $WS_2$.

## 4. PL and Raman data for MB and F4TCNQ-treated $MoS_2$

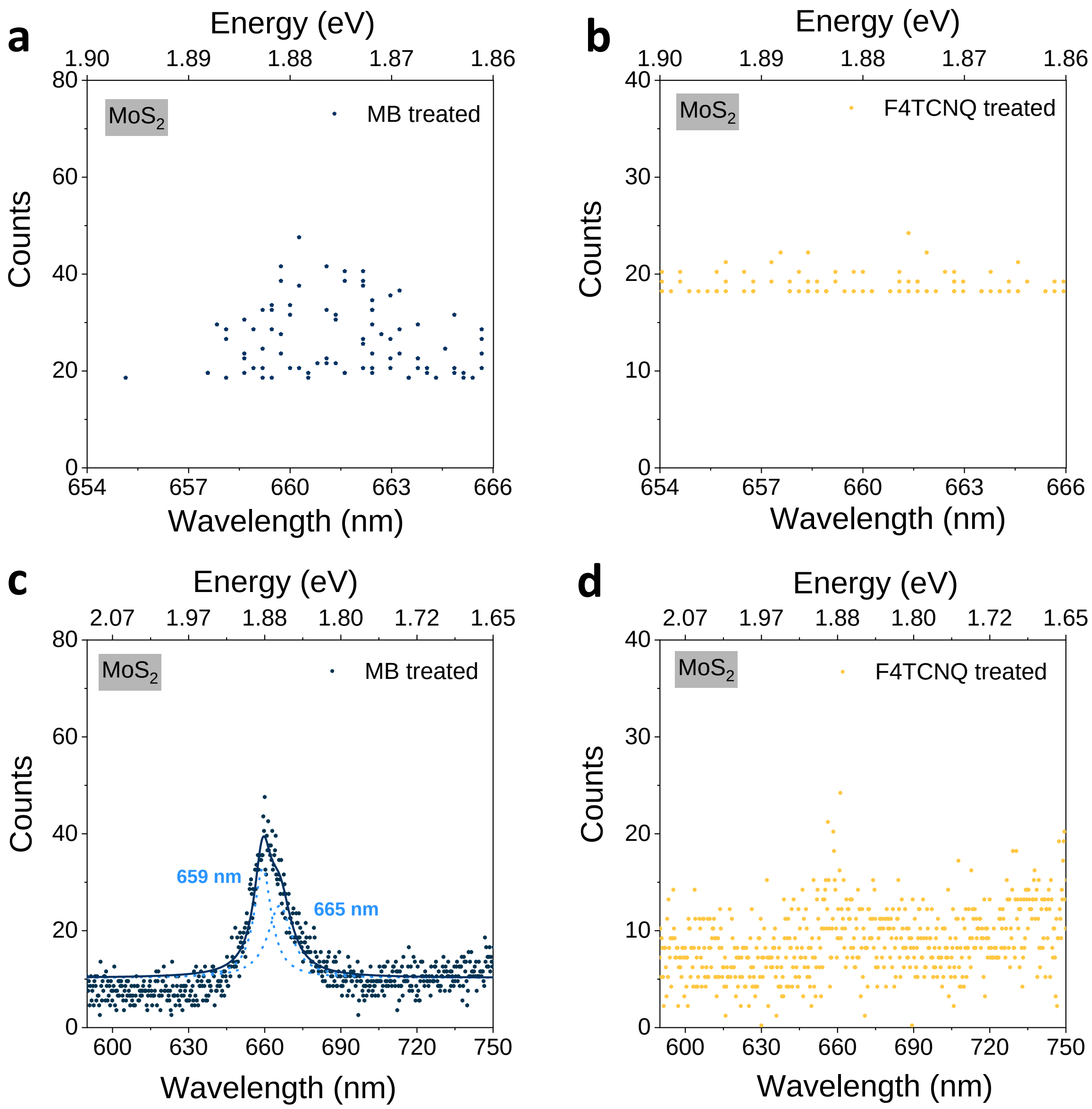

**Fig. S7 a** PL enhancement scatter plots showing peak MB-treated monolayer $MoS_2$ PL counts. **b** PL enhancement scatter plots showing peak F4TCNQ-treated monolayer $MoS_2$ PL counts. **c** Maximum PL spectrum for MB-treated monolayer $MoS_2$. **d** Maximum PL spectrum for F4TCNQ-treated monolayer $MoS_2$. The decomposed Lorentzian peak fitting of MB-treated $MoS_2$ is presented in dash line and the cumulative peak fitting is presented in solid line.

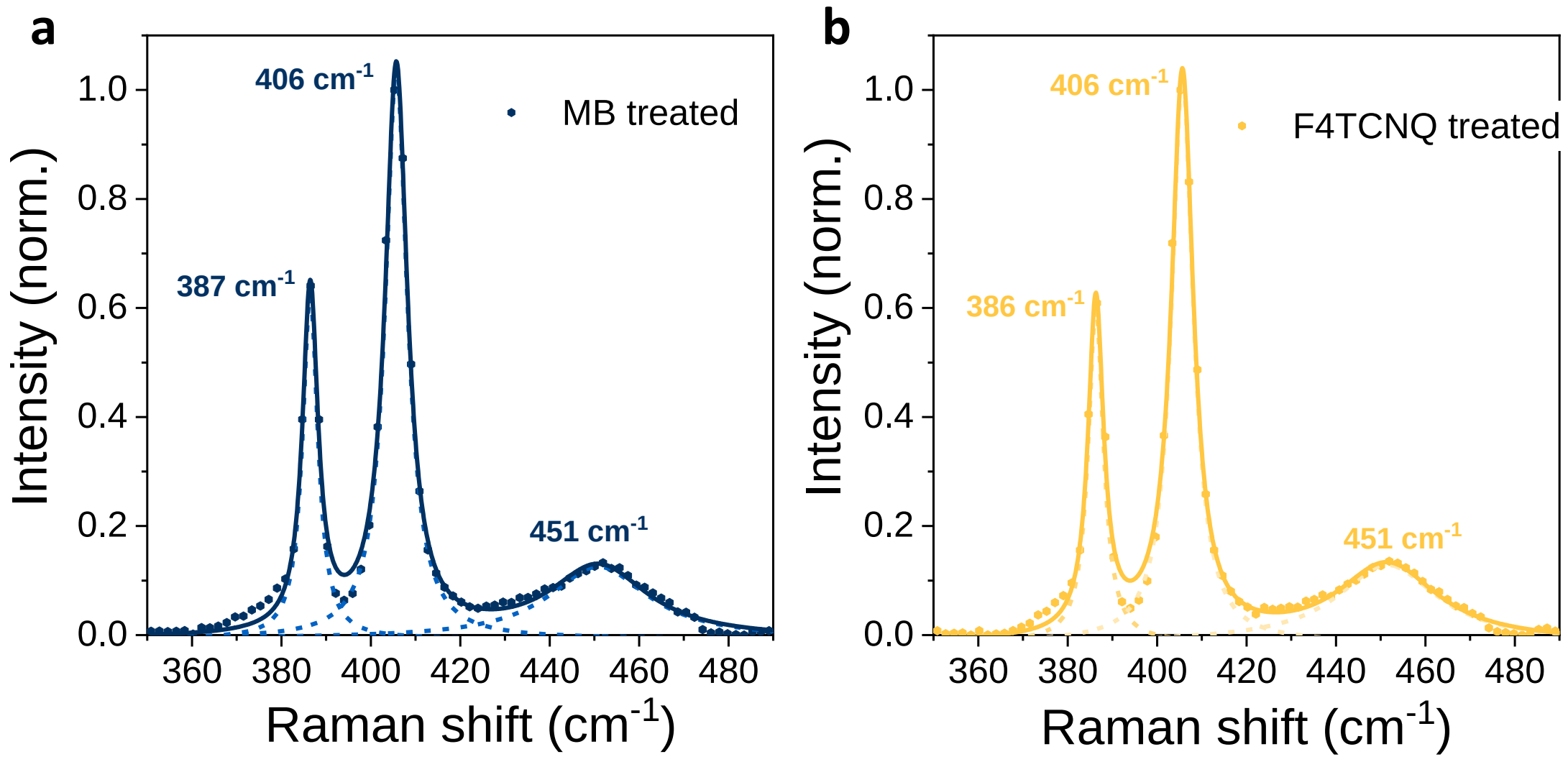

**Fig. S8** Raman spectra of **a** MB-treated, and **b** F4TCNQ-treated monolayer $MoS_2$. The decomposed Lorentzian peak fittings of MB and F4TCNQ-treated $MoS_2$ are presented in dash line and the cumulative peak fittings are presented in solid line.

## 5. XPS data for pristine H-TFSI and Li-TFSI treated $MoS_2$

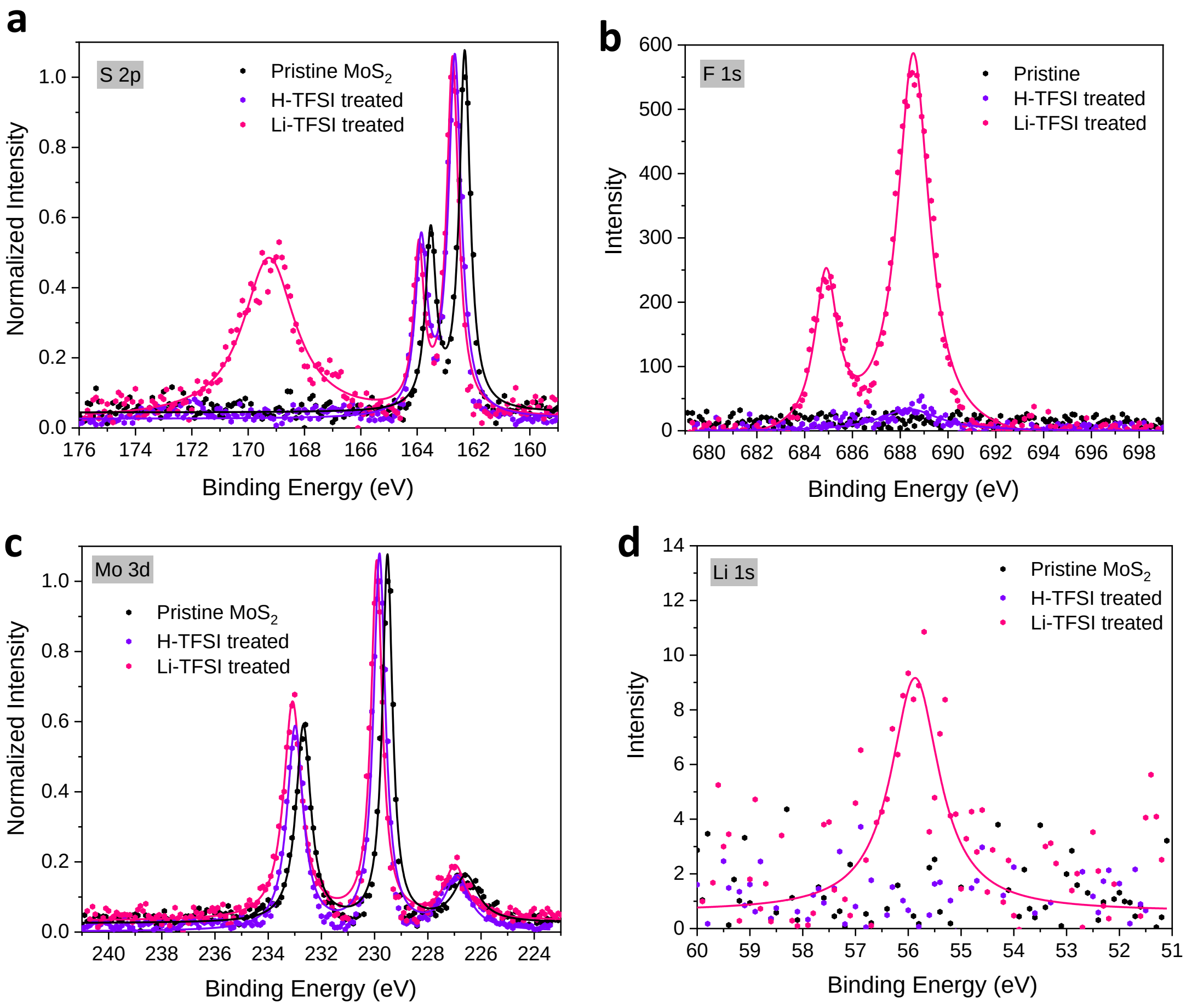

**Fig. S9** XPS spectra of pristine, H-TFSI-treated, and Li-TFSI-treated monolayer $MoS_2$. **a** Core level spectra of S 2p. **b** Core level spectra of F 1s. **c** Core level spectra of Mo 3d. **d** Core level spectra of Li 1s. The Lorentzian peak fittings of pristine and treated $MoS_2$ are presented are presented in solid lines.

## 6. Additional DFT simulation data for $WS_2$

**Table S1 a** DFT simulation of H and Li adsorption energies and the configurations on different positions of monolayer $WS_2$ surfaces.

| | $E^{Sv}$ (eV) | $E^{sf}$ (S) (eV) | $E^{sf}$ (Mo) (eV) |
|---|---|---|---|
| **H** | 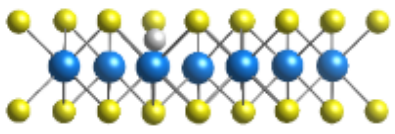 -2.29 | 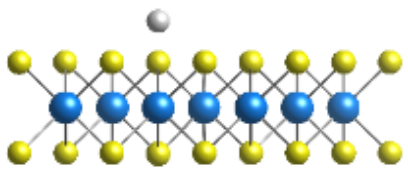 -0.08 | 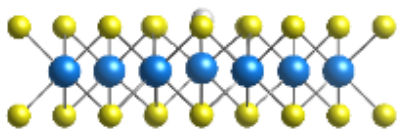 -0.27 |
| **Li** | 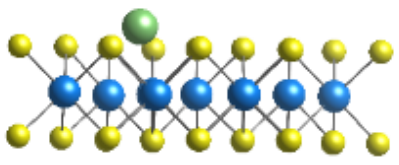 -2.12 | 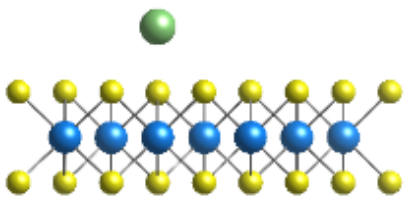 -0.65 | 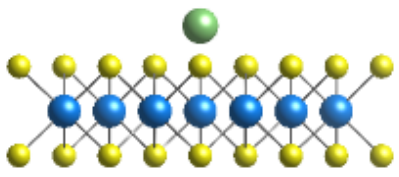 -1.33 |

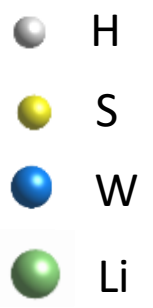

**Table S1 b** DFT simulation of Na, K, Ca and Mg adsorption energies and the configurations on different positions of monolayer $WS_2$ surfaces.

| | $E^{Sv}$ (eV) | $E^{sf}$ (S) (eV) | $E^{sf}$ (W) (eV) |
|---|---|---|---|
| **Na** | 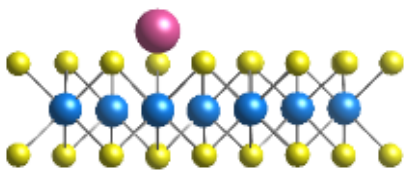 -1.72 | 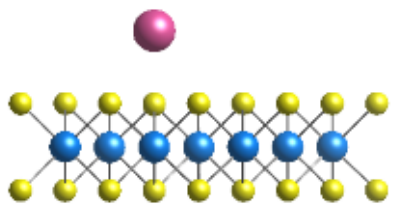 -0.47 | 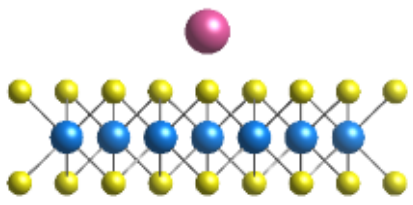 -0.78 |
| **K** | 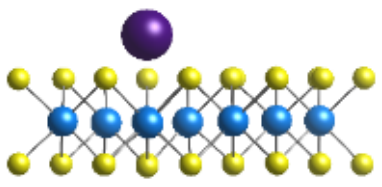 -2.02 | 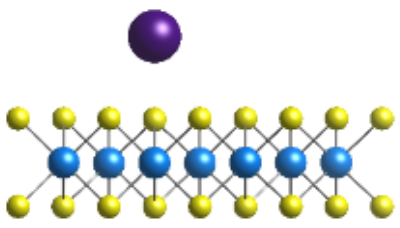 -0.76 | 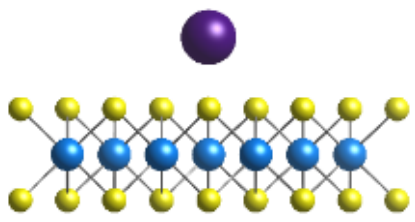 -1.02 |
| **Mg** | 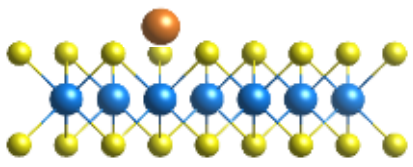 -0.90 | 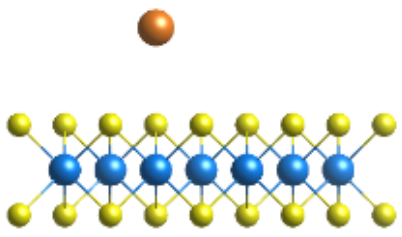 -0.06 | 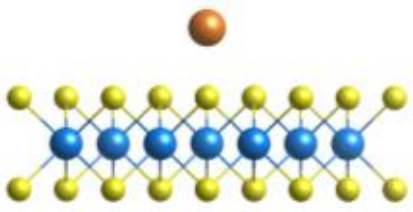 -0.15 |

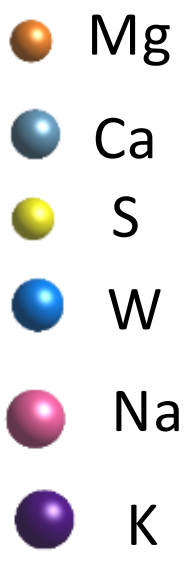

| | | | |
|---|---|---|---|
| **Ca** | 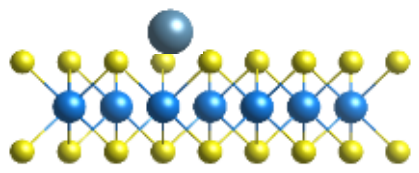 | 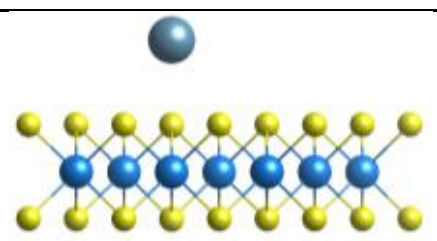 | 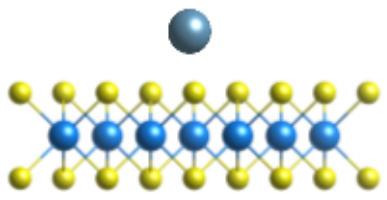 |
| | -2.83 | -0.37 | -0.80 |

**Table S1 c** DFT simulation of bond energies between $M_1^+$ and TFSI anion

| Bond | Bond Energy (eV) |
|---|---|
| H-TFSI | 4.73 |
| Li-TFSI | 5.37 |
| Na-TFSI | 4.74 |
| K-TFSI | 4.74 |

As shown in Table S1 a, b, similar with $MoS_2$, all adatoms on $WS_2$ present a clear preference of adsorption at sulphur vacancy sites compared to the surface of TMDSs. The adsorptions of Li adatom are generally energetically more favourable at surface sites compared to other $M_1$ atoms. Even though that the adsorptions of $M_1$ adatoms are energetically more favourable compare to $M_2$ atoms, each $M_2$ adatom contributes two positive charges, which explains the effectiveness of $M_2$TFSI treatments on improving the PL. On the other hands, the effectiveness of $M_1$-TFSI and $M_2$TFSI treatments on enhancing PL of TMDSs may also be related to how strongly the cations interact with TFSI anion. This determines the amount of cations interacting with the surfaces of monolayer TMDSs, therefore, the bond energy between cation and TFSI anion is simulated. As shown in Table S1 c, all cations present weak interactions with TFSI anion. Moreover, since the solution with ionic salts used during the chemical treatments is dilute and excessive, we assume there are enough cations interacting with the surface of TMDSs in all cases.

# 7. Pump-probe data for pristine, H-TFSI and Li-TFSI treated $MoS_2$

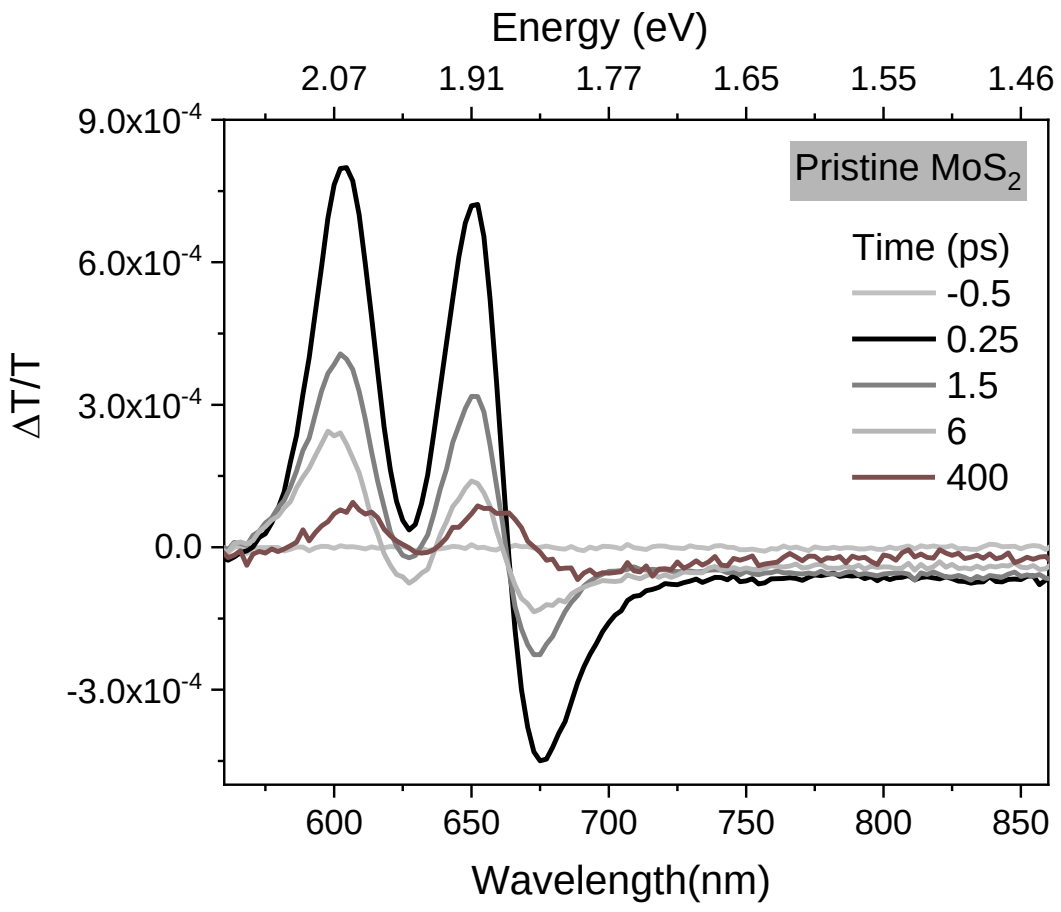

**Fig. S10** Pump-probe data of pristine $MoS_2$.

## 8. Additional PL data for $M_3$-Tf and Li-OAc-treated $MoS_2$ and $WS_2$

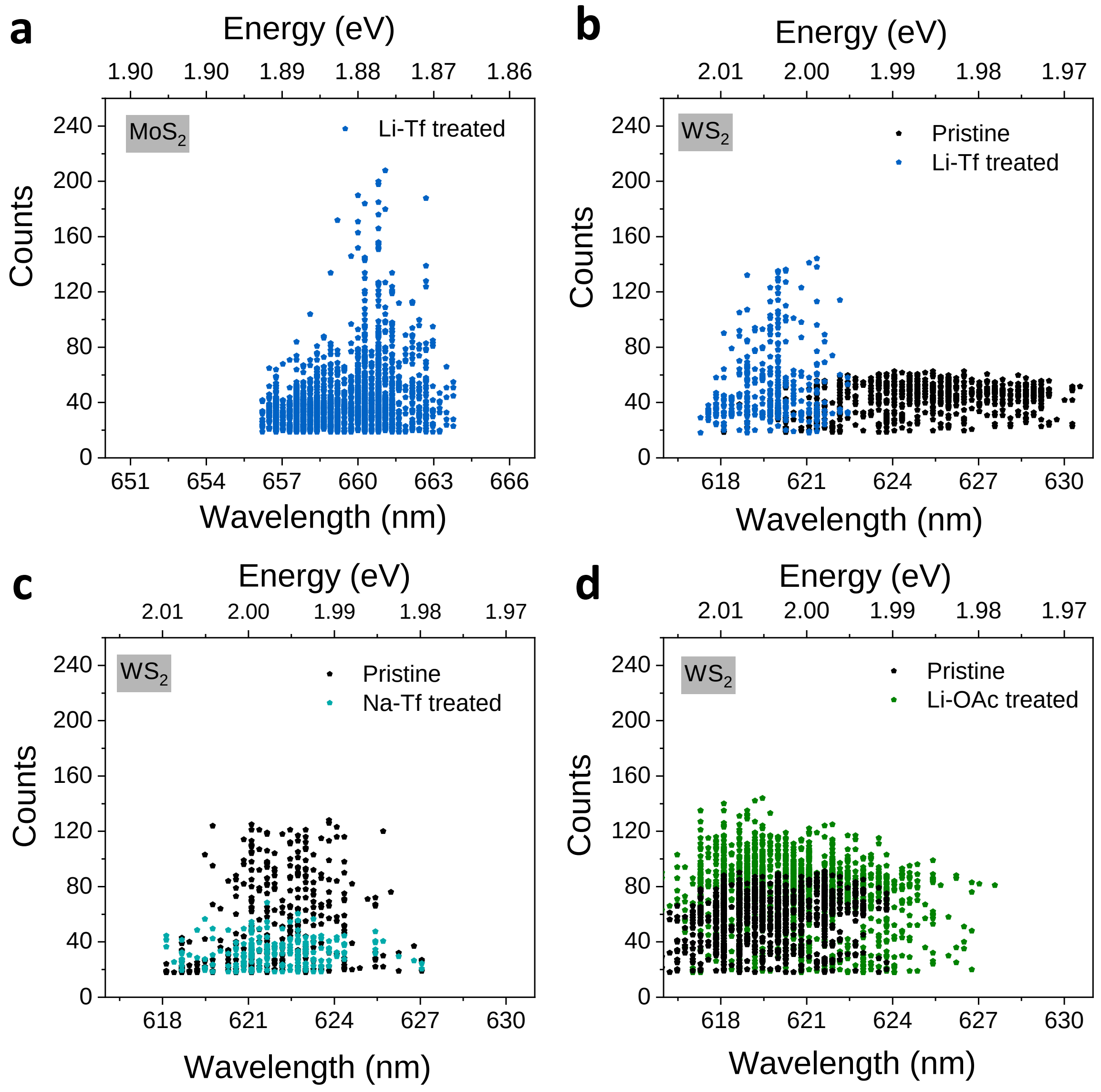

**Fig. S11** PL enhancement scatter plots showing peak **a** Li-Tf-treated monolayer $MoS_2$ PL counts, **b** pristine and Li-Tf-treated monolayer $WS_2$ PL counts, **c** pristine and Na-Tf-treated monolayer $WS_2$ PL counts, and **d** pristine and Li-OAc-treated monolayer $WS_2$ PL counts.

## 9. DFT simulation of anion adsorption on $MoS_2$ surface

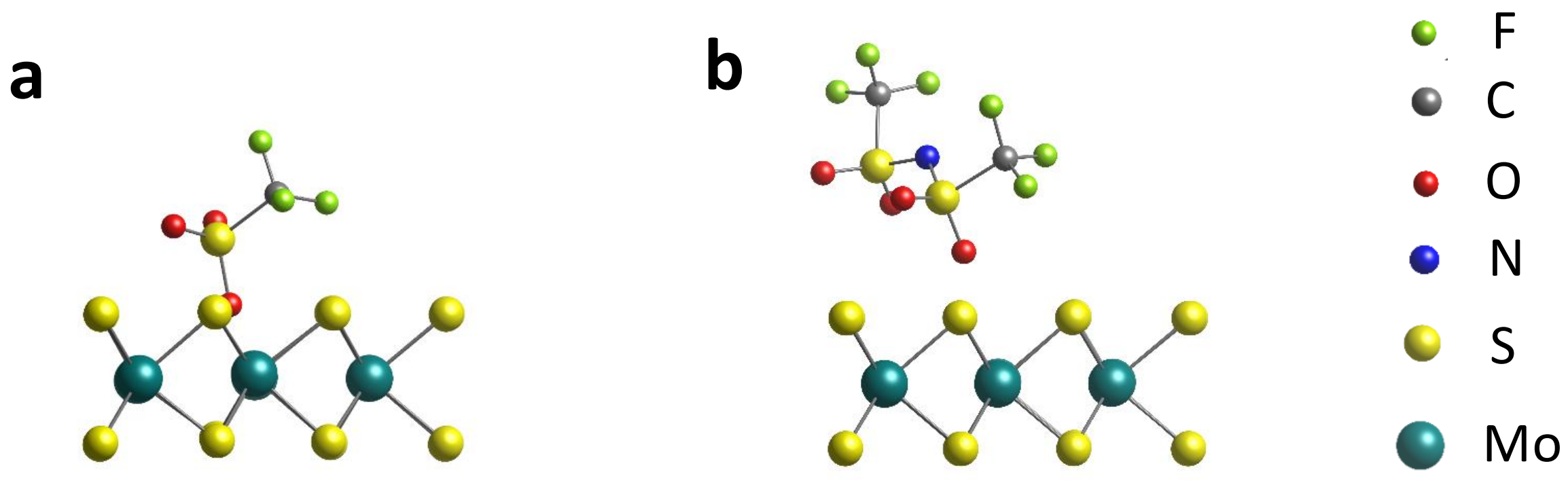

**Fig. S12** DFT simulation of **a** Tf and **b** TFSI anion adsorption at sulphur vacancy sites of monolayer $MoS_2$ surfaces.

## 10. Raman data for $M_3$-Tf and Li-OAc-treated $MoS_2$

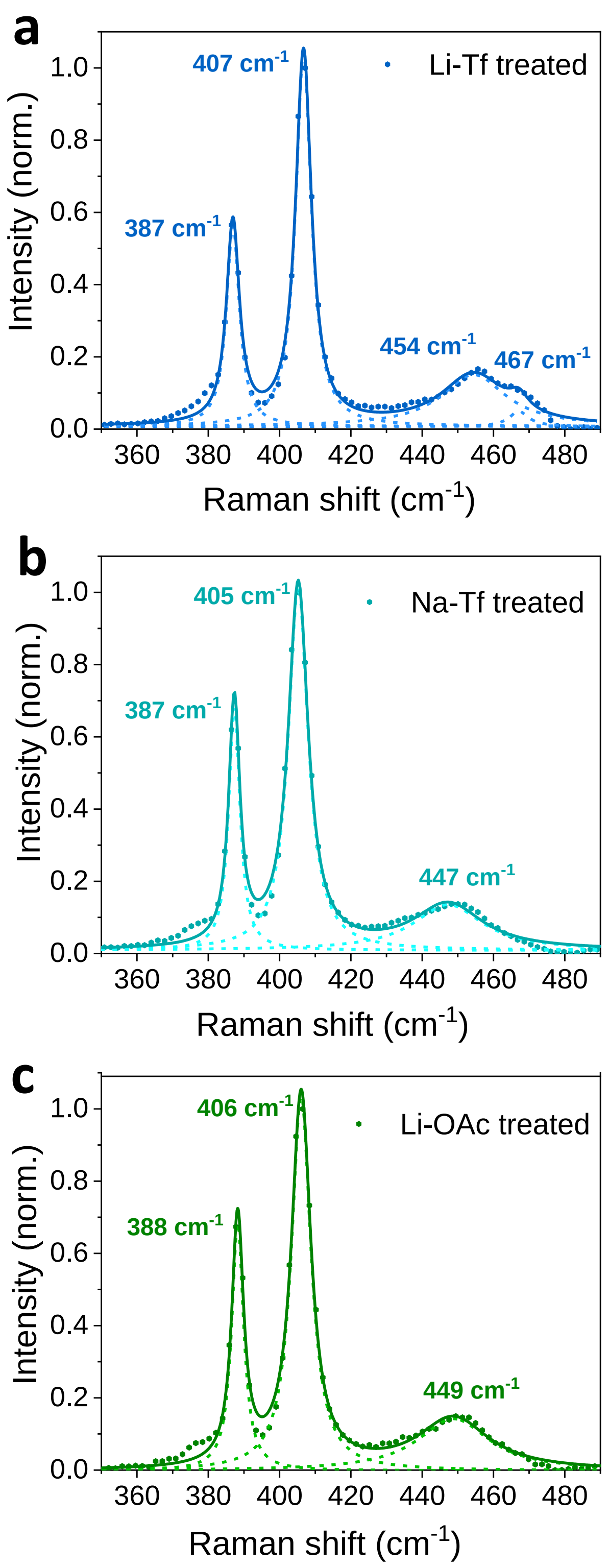

**Fig. S13** Raman spectra of **a** Li-Tf-treated, **b** Na-Tf-treated, and **c** Li-OAc-treated monolayer $MoS_2$. The decomposed Lorentzian peak fitting of each spectrum is presented in short dash line and the cumulative fitting is presented in solid line.

# 11. TRPL and PL diffusion data for Li-Tf-treated $MoS_2$

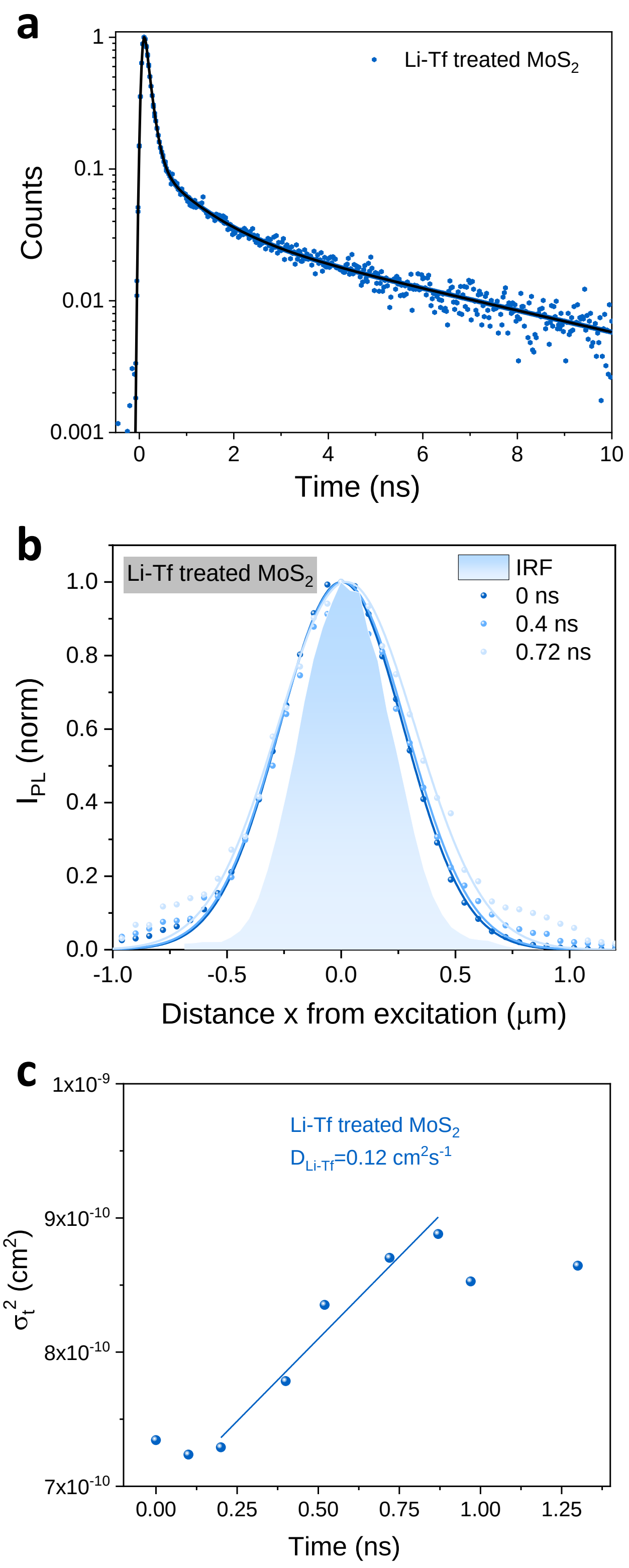

**Fig. S14 a** TRPL decay curve for Li-Tf-treated monolayer $MoS_2$. **b** Spatial profile of the normalized PL intensity $I_{PL}$ at snapshot $t = 0$, 0.4 and 0.72 ns for Li-Tf-treated monolayer $MoS_2$. **c** Corresponding $\sigma_t^2$ as a function of time.

# 12. Reference

1. Kash, J. A. Comment on ‘origin of the stokes shift: A geometrical model of exciton spectra in 2D semiconductors’. *Phys. Rev. Lett.* **71**, 1286 (1993).
2. Kolesnichenko, P. V., Zhang, Q., Zheng, C., Fuhrer, M. S. & Davis, J. A. Disentangling the effects of doping, strain and defects in monolayer WS2 by optical spectroscopy. (2019) doi:10.1088/2053-1583/ab626a.
3. Raja, A. *et al.* Dielectric disorder in two-dimensional materials. *Nat. Nanotechnol.* **14**, 832–837 (2019).
4. Amani, M. *et al.* Near-unity photoluminescence quantum yield in MoS2. *Science (80-. ).* **350**, 1065–1068 (2015).
5. Lien, D. H. *et al.* Electrical suppression of all nonradiative recombination pathways in monolayer semiconductors. *Science (80-. ).* **364**, 468–471 (2019).
6. Bertolazzi, S., Gobbi, M., Zhao, Y., Backes, C. & Samorì, P. Molecular chemistry approaches for tuning the properties of two-dimensional transition metal dichalcogenides. *Chem. Soc. Rev.* **47**, 6845–6888 (2018).
7. Rahm, M., Hoffmann, R. & Ashcroft, N. W. Atomic and Ionic Radii of Elements 1–96. *Chem. - A Eur. J.* **22**, 14625–14632 (2016).